# Diverse AI Personas Can Mitigate the Homogenization Effect in Human-AI Collaborative Ideation

Yun Wan
University of Houston - Downtown
wany@uhd.edu

Yoram M Kalman
The Open University of Israel
yoramka@openu.ac.il

## Abstract

Recent studies suggest that while generative AI (GenAI) can enhance individual creativity, it often reduces the diversity of collective outputs. A well-known example of this homogenization effect is by Doshi and Hauser (2024) who found that GenAI-generated plot ideas improved story writing creativity but led to convergence across writers' outputs. This study extends their experiment, identifying the design choices behind the apparent creativity-diversity trade-off. In Phase 1, we used structured prompting with 10 diverse GenAI personas to generate 300 story plots, and confirmed the plots' diversity using text embedding analysis. In Phase 2, participants wrote stories with or without access to these plots. Results show that diverse GenAI inputs can preserve story diversity compared to a human-only baseline, with some evidence of enhancement in the 1-plot condition. Beyond addressing the diversity component of the trade-off, our findings offer broader insights for human–AI system design. Our findings suggest that the trade-off may emerge from uniform deployment practices rather than from an inherent limitation of GenAI, and that diversity can be intentionally built into AI-mediated collaboration. Our study highlights the risks of over-standardization, the importance of prompt variation, and the value of treating GenAI not as a static tool but as a configurable partner. These insights have important implications for the design of GenAI systems that support, not constrain, collective creativity.

## 1 Introduction

The emergence of generative artificial intelligence (GenAI) is fundamentally transforming the landscape of creative work. Creativity is "the production of a novel and appropriate response, product, or solution to an open-ended task" (Amabile, 2013), and we are entering an era where AI systems do more than simply support human creativity: they actively participate in co-creative processes. This represents what researchers term "Human-AI Co-Creativity", referring to human users collaborating with an AI agent on a shared creative task in real time (Karimi et al., 2020; Wan et al., 2024). Rather than automating creativity, the real promise of AI in this context lies in how it can amplify and support human creative potential. Recent advances in large language models and multimodal AI systems have enabled sophisticated forms of creative collaboration that were previously unimaginable. Some creative tasks can now be fully automated, which becomes evident, for example, with the generative fill function in Photoshop, code generation in IT, or character design in video games. However, the most promising applications emerge when AI serves not as a replacement but as an active creative partner, contributing unique capabilities that complement human creativity (Sako, 2024).

The theoretical foundations of such collaboration draw from decades of research on human teamwork and collective intelligence (Salas et al., 2008; Suran et al., 2020). The diversity of participants positively affects collective problem-solving performance and may in some cases be more important than their individual abilities, with actor diversity becoming progressively more important as problems grow in complexity and require exploring novel solutions (Page, 2008). This principle has inspired organizations across many sectors to adopt cross-functional teams and now presents an opportunity to extend this approach to human-AI teams. Contemporary research reveals that effective human-AI collaboration operates across multiple levels of sophistication (Kolbjørnsrud, 2024; Kong et al., 2025; Przegalinska et al., 2025; Wu et al., 2025). Firms can deploy AI to automate routine tasks and enhance decision support tasks but should focus on AI and human collaboration to boost creative and innovative tasks (Przegalinska et al., 2025; Wu et al., 2025). At the highest level of collaboration, both entities contribute significantly to a shared creative product, where the inputs and outputs of humans and AI blend seamlessly, resulting in a synergistic creative process that transcends traditional boundaries (Gianet et al., 2024; Kong et al., 2025; McGuire et al., 2024).

Yet despite these promising developments, a frequently encountered challenge has emerged that threatens to limit the potential of human-AI creative collaboration: *the creativity-diversity trade-off*. Following Stirling's (2007) integrative framework, we use diversity to mean the pattern of differences within a collection along three necessary properties—variety (how many kinds are present), balance (how evenly those kinds are represented), and disparity (how far apart those kinds are). In our context, this encompasses both human diversity and textual diversity.

Recent studies on the contributions of GenAI to human performance of creative tasks demonstrated this trade-off: GenAI can boost the quality of the creative outputs, but this involves a reduction in the diversity of the outputs (e.g., Chen & Chan, 2024; Dell'Acqua et al., 2023; Doshi & Hauser, 2024; Veselovsky et al., 2023).

The most prominent of these studies is Doshi and Hauser's (2024) paper which demonstrated that GenAI-generated plot ideas enhanced the creativity of individual storywriters but, notably, reduced the collective diversity of their stories. The experiment showed that when participants had access to GenAI-generated plot ideas, their stories were rated as more creative on an individual level, but as a collective, their stories were more similar to each other than stories written by participants who did not have access to the plot ideas. This was evidenced by the cosine similarity scores rising by 10.7% and 8.9% for participants with access to one or five GenAI plot ideas, respectively. In another study Boston Consulting Group consultants using GenAI were found to generate more solutions to client problems and these solutions were rated as higher quality, but according to their semantic similarity metrics, the diversity of solutions decreased (Dell'Acqua et al., 2023). In still another study, Chen and Chan (2024) demonstrated that human users using GenAI to ghostwrite or review ad copies produced ads with higher linguistic quality and effectiveness, but the semantic similarity of user output increased by 60% and 32%, respectively, indicating a significant standardization effect. Veselovsky et al. (2023) also reported this standardization effect in a study of crowd workers summarizing scientific abstracts. They found that the summaries created with LLM assistance preserved more keywords from the original texts (40.1% vs 31.2% for human-only summaries) but were significantly more homogeneous, with homogeneity scores of 45.6% for LLM-assisted texts compared to just 27.1% for human-generated texts, demonstrating how GenAI tools can simultaneously improve quality while reducing the diversity of collective outputs.

All these findings lead to the question: *Is the trade-off between GenAI assistance in a creative task and the diversity of the creative outcome an inevitable result of working with GenAI?* We hypothesize that it is not. This hypothesis is currently supported by a small number of studies that attempted to tackle the homogenization effect of GenAI assisted ideation: Meincke et al. (2024) demonstrated that prompt engineering can substantially improve the diversity of AI generated ideas, and of these, Chain-of-Thought (CoT) prompting was the most effective prompting approach; Bordas et al. (2025) focused on the well-studied fixation effect in the creativity literature and demonstrated that diversity can be increased by using design-based de-fixation techniques, most effectively introducing new, unrelated knowledge to the ideation process.

Indeed, we too propose that the reduced diversity observed in human-AI collaborative outcomes stems not from an inherent limitation of GenAI technology. Rather, we suggest that the reduced diversity in outcomes stems from the similarity of ideas participants received from GenAI. Consider, for instance, a popular interior design tip: placing mirrors opposite windows to enhance natural light. When many designers adopt this suggestion, the resulting spaces may indeed appear brighter and more attractive—but they also begin to look increasingly alike. A parallel can be seen in marketing, where a widely shared tip for writing high-performing headlines may boost click-through rates (at least initially), yet also leads to headlines that follow the same predictable patterns. In both cases, individual performance improves, but variation across outputs diminishes.

We demonstrate how the use of homogeneous prompting strategies fails to leverage the full generative potential of GenAI systems by closely examining the experimental design of Doshi and Hauser (2024). In their study, all participants who received GenAI assistance were exposed to plot ideas generated from identical prompts. While each individual participant benefited from exposure to AI-generated ideas that were novel *to them*, a similar set of AI-generated plot ideas was presented to *all* participants. This creates a convergence effect: when different humans are exposed to highly similar external creative stimuli, their outputs naturally gravitate toward similar conceptual territories, thereby reducing collective diversity despite individual quality improvements. The underlying logic is straightforward yet profound: if GenAI provides ideas that are genuinely new to each user, their individual work will improve on average, precisely because the AI's perspective differs from each individual's baseline creativity. However, when this same "novelty" is uniformly distributed across different participants, it creates a shared reference point that constrains the collective creative space. The diversity reduction is therefore not a consequence of Human-AI collaboration per se, but rather an unintended consequence of design decisions that limit the AI's contribution to a narrower range of outputs.

An initial test of this proposition was replicating the method Doshi and Hauser (2024) used to supply the participants with GenAI-generated plot ideas. We provided the same prompt to the same OpenAI GPT-4 API, and received plot ideas that were unique. Yet, even before we applied any quantitative tools, it was plain to see they were highly similar to each other. This part of the experiment is described in more detail later in the method section of Phase 1, and in the Supplementary Materials.

To test the hypothesis that the source of story diversity (or lack thereof) is in the diversity of ideas the GenAI provides, we extended the core experiment from Doshi and Hauser (2024) with one key modification: instead of using a uniform prompt to elicit Gen-AI generated plot ideas, we intentionally diversified the GenAI plot idea outputs available to human participants. We achieved this diversification by generating a heterogeneous idea pool through systematic prompting of 10 distinct GenAI "personas," each designed to embody different creative perspectives, cultural backgrounds, and narrative approaches among other dimensions (Ge et al., 2024; Park et al., 2024). This methodological innovation allows us to maintain the benefits of human-AI collaboration while preserving – and potentially enhancing – the diversity of collective creative outputs.

The theoretical foundation for our persona-based approach draws directly from established research on team diversity and collaborative creativity. Diverse teams, characterized by varied cognitive styles, backgrounds, and perspectives, can produce more innovative and creative results than homogeneous groups (Aggarwal & Woolley, 2019; Freeman & Huang, 2014). The critical insight is that previous studies focused on GenAI-generated creative ideas, without actively encouraging the diversity of these generated ideas. All participants in these experiments essentially collaborated with identical AI partners, despite working on individual creative tasks. Our persona-based approach transforms this dynamic by ensuring that participants collaborate with distinctly different AI perspectives, analogous to assembling a diverse creative team rather than asking the same collaborator again and again for a new idea.

Our experimental implementation created 10 distinct GenAI personas, each with unique creative orientations and perspectives. While individual personas generated internally consistent (and therefore less diverse) plot ideas, the increased variation across personas was statistically significant, providing a sufficiently heterogeneous input pool necessary for maintaining collective diversity. The results validated our hypothesis: when human participants wrote stories using plot ideas from diverse AI personas, their collective outputs maintained the same level of diversity as stories written without any GenAI assistance, and even exceeded it. This finding mitigates the homogenization effect.

In the following sections, we present the empirical validation of this theoretical framework through a comprehensive two-phase experimental design. Phase 1 establishes the diversity characteristics of our persona-generated plot ideas, while Phase 2 examines their impact on the diversity of human creative outputs.

# 2 Phase 1: AI Plot Diversification

## 2.1 Phase 1 Method

We developed 10 distinct GenAI personas using ChatGPT's o1-preview model through a systematic, multi-stage process designed to maximize diversity across multiple dimensions. The persona creation methodology included three sequential phases: identifying relevant dimensions of variation, expanding and refining a set of trait categories, and generating detailed personas. We began by prompting the model to suggest a set of effective dimensions along which creative personas might differ. This meta-prompting approach yielded a set of candidate dimensions including cultural background, literary genre preferences, and cognitive thinking styles (see Supplementary Materials for the full list).

Building on these initial dimensions, we expanded the framework into 14 categories, each containing between 4 and 20 distinct trait classes. These included, for example, cultural background (e.g., East Asian, Indigenous Cultures), cognitive styles (e.g., analytical, divergent thinking), professional interests (e.g., gardening, digital media), and literary preferences. We explicitly excluded literary genre preferences related to adventure in order to avoid biasing personas toward the genre participants would later engage with (see Supplementary Materials for the complete taxonomy). Using this framework, we then prompted the model to generate 10 detailed persona profiles of at least 100 words each. Each persona combined randomly sampled traits from across the categories to ensure systematic variation across the multidimensional space. These profiles served as the foundation for all subsequent plot generation (see Supplementary Materials for all persona descriptions).

The personas were then deployed in a custom assistant built on the Azure OpenAI platform, using GPT-4o (version 2024-05-13, temperature = 0.9, top P = 1). Each persona was instructed to generate thirty-one plot ideas, each consisting of exactly three sentences, using the prompt: "Create thirty-one plot ideas of exactly three sentences each that [persona name] would write." For each persona, the first 30 plots were selected and retained. To ensure sufficient variety in the generated outputs, ChatGPT was used to identify cases where multiple plots were substantively similar. Early runs at a temperature setting of 0.7 yielded several redundant plots, but this issue was resolved by increasing the temperature to 0.9 (see Supplementary Materials for the full prompt and sample plot outputs)

An additional comparative dataset was created, as mentioned briefly in the Introduction, of 30 plot ideas created through the same means used by Doshi and Hauser (2024). The prompt "write a three-sentence summary of a story about an adventure on a different planet" was run 30 times on the Azure OpenAI chat playground using GPT-4, to emulate running this prompt on the OpenAI API used by Doshi and Hauser (2024). The 30 plot ideas were collected.

## 2.2 Analyses of the AI persona plot ideas

In Phase 1, to establish this foundational premise of our intervention—that diverse AI personas can generate meaningfully different creative content—we conducted a comprehensive similarity analysis of the 300 story plots generated by our 10 distinct personas (30 plots per persona). We examined two key measures: variation within individual personas (intra-persona similarity) and variation across different personas (inter-persona diversity). This analysis employed cosine similarity calculations on text embeddings to quantify the similarity within individual personas (intra-persona similarity) and between different personas (inter-persona similarity).

Our analysis used a systematic pairwise comparison approach to measure cosine similarity differences between stories with adequate statistical power. We calculated similarity in two ways: within the plots created by the same persona and between plots created by different personas.

For within-persona similarity, we compared every possible pair of the 30 stories generated by each persona. This yielded 435 unique comparisons per persona (calculated as C(30,2) = 435), totaling 4,350 within-persona comparisons across all 10 personas. For between-persona similarity, we compared stories across different personas. Each story from one persona was compared to every story from each of the other nine personas. This resulted in 900 comparisons between each pair of personas (calculated as 30 × 30 = 900). Across all possible combinations of the 10 personas, we conducted 40,500 total between-persona comparisons. This number comes from comparing each of the 45 possible persona pairs (calculated as C(10,2) = 45), with each pair generating 900 comparisons.

This approach captures both how consistent each persona's stories are internally and how different the personas are from each other and provides strong empirical evidence for measuring the distinctiveness of each persona's creative output.

We also calculated each persona's distinctiveness score to quantify how consistently each AI persona generates similar content within itself compared to other personas. The distinctiveness score equals the average intra-persona similarity minus the average inter-persona similarity for each persona. Positive scores indicate that a persona's story plots are more similar to each other than to story plots from different personas, suggesting a coherent creative "voice." Higher positive scores reflect greater persona distinctiveness. This metric helps identify which AI personas have developed the most consistent and recognizable narrative styles.

All similarity calculations employed OpenAI's "text-embedding-ada-002" model, which generates 1,536-dimensional vectors that provide high-resolution semantic representations of plot content. These embeddings capture subtle thematic, structural, and stylistic differences in narrative content, enabling precise quantification of semantic relationships between story plots. The cosine similarity metric, ranging from 0 (completely dissimilar) to 1 (identical), was applied to these embedding vectors to generate similarity coefficients for all pairwise comparisons. This analysis design enables direct statistical comparison between intra-persona and inter-persona similarity distributions.

## 2.3 Phase 1 Results and Discussion

### 2.3.1 Story plots generated by GenAI personas are more diverse

The analysis of the AI persona plot ideas is presented in Table 1. All cosine similarity scores are scaled to 100 here and in subsequent reports.

***Table 1***

*Summary Statistics of Intra- and Inter-Persona Similarities*

| | Result |
|---|---|
| Mean intra-persona similarity | 87.41 ± 3.17 |
| Mean inter-persona similarity | 82.93 ± 3.12 |
| Difference | 4.49 |
| T-test | t = 119.9443, p < 0.00000001 |
| Effect size (Cohen's d) | 1.4254 |
| Most distinctive persona | Persona 6 (score: 7.69) |
| Least distinctive persona | Persona 7 (score: 2.48) |

Furthermore, the same analysis of the 30 plot ideas created by the OpenAI API using the prompt used by Doshi and Hauser (2024) yielded the highest average intra-group similarity: 90.99 (SD = 2.78), confirming our sense that the plot ideas generated by directly prompting the generic model lead to the least diversified, in other words most homogenous, story plots.

Our analysis demonstrates robust persona distinctiveness in AI persona-generated plots, with mean intra-persona similarity of 87.41 (SD = 3.17) significantly exceeding inter-persona similarity of 82.93 (SD = 3.12). This difference of 4.49, while numerically modest, proves statistically compelling with t = 119.9443 (p < 0.00000001) and represents a very large effect size (Cohen's d = 1.4254), establishing that our persona-based system generates content with meaningful differentiation between different personas' plots - a finding that demonstrates substantial practical significance for computational narrative generation.

### 2.3.2 Interpretation of High Baseline Similarities

The relatively high baseline similarities (both intra and inter-persona) merit careful interpretation within the context of AI-generated creative content. These high values (87 and 83 respectively) reflect several important characteristics of our experimental design and the nature of generative AI systems.

First, all personas were constrained to generate content within the same genre (adventure stories set on other planets) and format (exactly three sentences), creating structural similarities that elevate baseline similarity scores. This constraint was intentional, designed to test whether personas could achieve meaningful differentiation even within tightly controlled creative parameters – a more stringent test of persona effectiveness than would be provided by allowing unlimited creative freedom.

Second, the high similarities reflect the shared statistical foundations of large language models, which are trained on overlapping datasets and optimized for coherent, contextually appropriate text generation. The fact that our personas achieved statistically significant and practically meaningful differentiation despite these shared foundations provides evidence for the effectiveness of persona-based approaches.

Third, the adventure genre itself contains recognizable structural and thematic elements (protagonists, challenges, resolutions) that naturally increase semantic similarity across all generated content. The persona approach succeeded in creating differentiation within these genre constraints, suggesting that similar approaches could achieve even greater differentiation in less constrained creative domains.

#### 2.3.3 Individual Persona Performance Analysis

Analysis of individual persona distinctiveness reveals interesting variation in the effectiveness of different persona configurations, providing insights into which characteristics more successfully generate unique creative content. For example, Persona 6, characterized as *Carlos from the Caribbean* with a passion for comedy, divergent thinking style, and technophile orientation, achieved the highest distinctiveness score of 7.69. This persona's plots consistently demonstrated unique characteristics that set them apart from other personas' outputs. Examination of Persona 6's content reveals several distinguishing features that likely contribute to its high distinctiveness:

The comedy orientation appears to be a particularly powerful differentiating factor, as humorous elements in adventure plots create distinctive semantic signatures that are readily detected by embedding models. Plots generated by Persona 6 frequently incorporated elements of absurdity, wordplay, and incongruous situations that distinguish them from the more serious adventure narratives produced by other personas. For example, where other personas might describe "*treacherous alien terrain*," Persona 6 generated descriptions of "*bouncing like lunatics on air-filled streets*" and "*floating hotel drifting away into the cosmos*."

The divergent thinking style might also have contributed to Persona 6's distinctiveness by generating unexpected plot developments and creative solutions that deviate from conventional adventure narrative structures. This persona frequently introduced novel problems and solutions that other personas did not generate, such as characters needing to learn alien communication through "*inflatable balloon animals*" or dealing with technology malfunctions caused by "*candy data overloads*."

Persona 7, characterized as *Elena from North America* with graphic novel interests, holistic thinking style, and entrepreneurial orientation, achieved the lowest distinctiveness score of 2.48. While still demonstrating positive distinctiveness (indicating that intra-persona similarity exceeded inter-persona similarity), this persona's plots were more similar to those generated by other personas than any other in our system.

Analysis of Persona 7 for traits that make it less distinctive are obviously less straightforward. Yet one possible element is her North American cultural background and entrepreneurial orientation which may align closely with prevalent patterns in the training data of large language models, leading to outputs that cluster around statistical centroids rather than exploring more distinctive creative territories. This suggests that personas incorporating less common cultural perspectives or thinking styles may achieve greater distinctiveness in AI-generated content.

The full information about similarity and distinctive score for all personas are available in Table S1 In the Supplementary Materials.

#### 2.3.4 Distribution of Persona Distinctiveness

The remaining eight personas achieved distinctiveness scores distributed between these extremes. With scores ranging from 2.48 to 7.69, this distribution demonstrates that while all personas achieved meaningful differentiation, there is substantial room for optimization in persona design. The threefold difference between the most and least distinctive personas (7.69/2.48 = 3.10) indicates that persona characteristics significantly influence the uniqueness of generated content.

Figure 1 presents the overall distribution of similarity scores, revealing the clear separation between intra-persona and inter-persona similarity distributions. The minimal overlap between these distributions visually demonstrates the significance of our findings while illustrating the consistency of persona effects across all generated content. It also displays the persona-by-persona similarity (red line box), revealing systematic patterns in which personas are most and least similar to each other. This heatmap visualization demonstrates that persona similarities are not randomly distributed but follow patterns related to shared characteristics. An obvious example is persona 6 which stands out with relatively low (mainly blue) inter-persona similarity, yet relatively high (mainly yellow-green) intra-persona similarity.

***Figure 1***

*A heatmap of cosine similarity score (x100) between each of the 300 story plots and all the other plots. The red lines enclose plots generated by the same persona.*

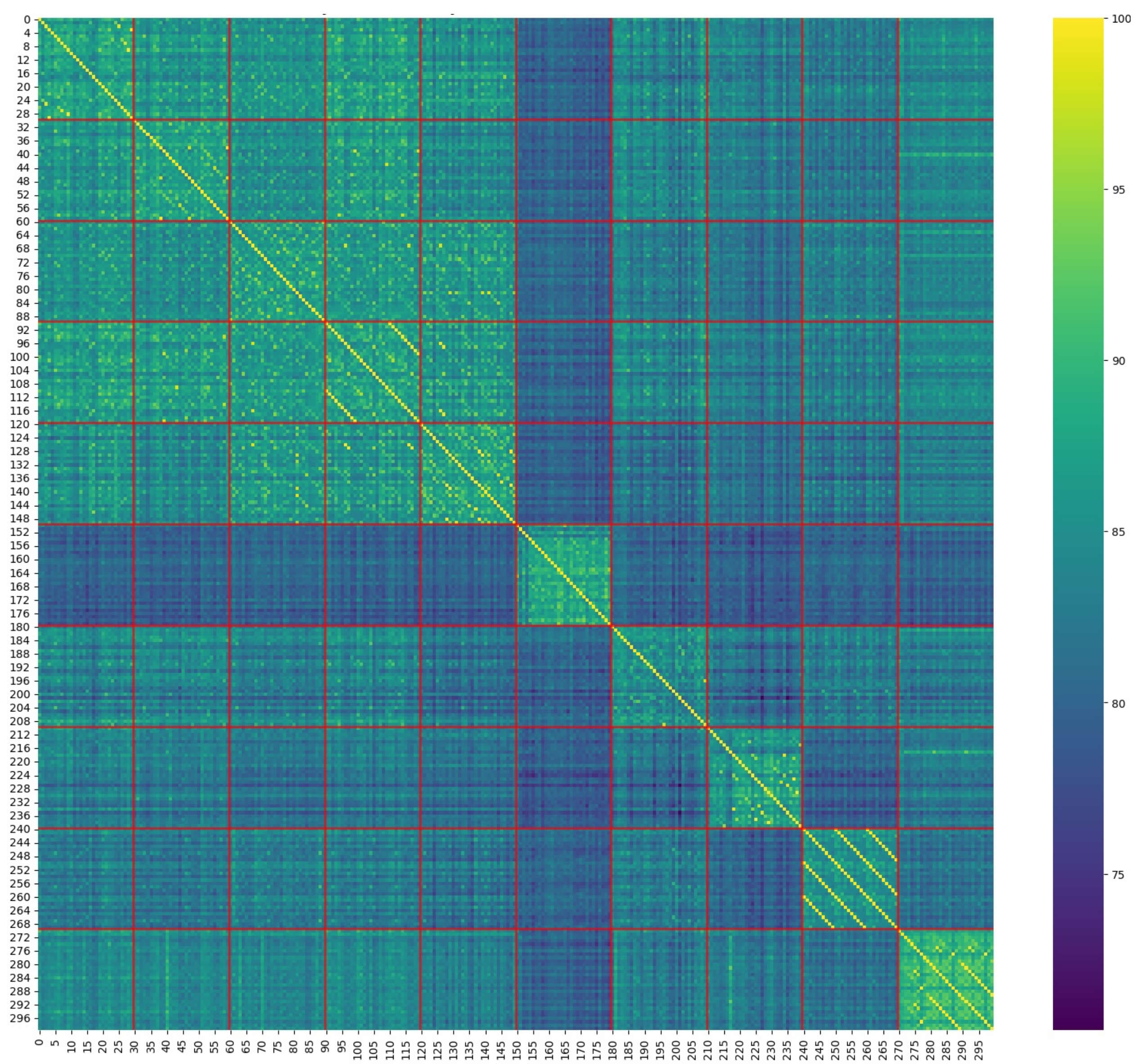

# 3 Phase 2: Human Story Creation Inspired by AI Plots

## 3.1 Phase 2 Method

After measuring the diversity of plot ideas generated by different GenAI personas, we recruited 100 human participants using the online platform Prolific, inviting UK-based participants with an approval rating at or above 95% and between 100 and 1,000,000 prior submissions (same criteria as Doshi and Hauser (2024)). The participants were randomly assigned to write an eight-sentence short story under one of three experimental conditions: a control group without AI support (0-plot control condition), a group with the option to request access to one AI-generated plot idea (1-plot condition), and a group with the option to request access to up to five AI-generated plot ideas (5-plot condition). See the Supplementary Materials for screenshots of the participants' experimental interface.

In the two AI-assisted conditions, participants received ideas randomly chosen from the diverse pool of 300 plot ideas generated by the 10 personas. After excluding data from three participants in the control group who reported using external AI to inspire their stories, we analyzed 97 stories: 32 in the control group, 33 in the 1-plot condition, and 32 in the 5-plot condition.

## 3.2 Analyses of the Short Stories

Following the same embedding approach used in Phase 1, we analyzed story diversity using cosine similarity calculations on OpenAI's 1,536-dimensional text embeddings. We employed three complementary analytical approaches to comprehensively assess the impact of diverse AI personas on collective story diversity.

*Replication Analysis*: *Similarity to Group Centroid.* To directly replicate the methodology of Doshi and Hauser (2024), we calculated cosine similarity scores between each individual story and the collective average (centroid) of all stories within the same experimental condition. This approach measures how closely individual stories conform to their group's average narrative profile, with higher similarity indicating greater convergence toward a common narrative template. We analyzed centroid similarity using linear regression to test for significant differences across conditions, with the control group as the reference category. Additionally, we conducted one-way ANOVA to compare mean centroid similarities across all three conditions. This dual analytical approach allows direct comparison with Doshi and Hauser's (2024) findings while testing our hypothesis that diverse AI personas can eliminate the homogenization effect.

*Mixed-effect Modeling:* Mixed-effects modeling accounts for the non-independence inherent in pairwise similarity comparisons by incorporating random effects for both stories in each pairwise comparison, while treating experimental condition as a fixed effect. This approach models the variance at two levels: between-story variance (captured by random effects) and within-condition variance in pairwise similarities. This formulation allows us to test whether experimental conditions significantly influence pairwise similarity while properly accounting for the fact that the same stories contribute to multiple observations in the dataset. We used restricted maximum likelihood (REML) estimation to fit the model, which provides unbiased estimates of variance components in unbalanced designs. As a robustness check, we complemented the mixed-effects analysis with a story-level averaging approach. For each story, we calculated its average similarity to all other stories within the same condition, creating one independent observation per story. We then conducted one-way ANOVA on these story-level averages to test for condition effects. This alternative approach sacrifices information about the distribution of pairwise similarities but ensures complete independence of observations, providing a conservative validation of the mixed-effects results.

*Linguistic Diversity Analysis: Traditional Text Analytics.* To examine narrative diversity beyond semantic embeddings, we conducted fine-grained linguistic analysis using the Natural Language

Toolkit (NLTK) (Bird et al., 2009) . Our text processing pipeline employed NLTK's sentence and word tokenizers for segmentation, followed by part-of-speech tagging to identify narrative elements including actions (verbs), descriptions (adjectives), and character names (proper nouns). We systematically quantified linguistic diversity by analyzing the variety of descriptive and emotional vocabulary across conditions, comparing AI-assisted stories (1-plot and 5-plot conditions combined) against human-only stories (control condition) using independent sample t-tests. This approach captures surface-level linguistic richness that may not be fully reflected in high-level semantic similarity measures.

This comprehensive analytical framework directly tests our central hypothesis: that diverse AI personas can maintain collective story diversity, effectively eliminating the homogenization effect observed in previous research.

## 3.3 Phase 2 Results and Discussion

### 3.3.1 Collective diversity of stories by GenAI-aided human participants increased

The replication analysis results indicated no significant difference in story similarity across conditions (1-plot condition: b = -0.493, p = 0.301; 5-plot condition: b = -0.356, p = 0.459, see Table 2). There was also no significant difference in the means (see Table 3). Most importantly, our linear regression in Table 2 does not show the statistically significant negative impact of the access to GenAI ideas on the similarity of stories within the same experimental condition that reported in Doshi and Hauser (2024), nor are the means in Table 3 statistically significantly different from each other ($F(2, 94) = 0.573$, $p = 0.57$).

***Table 2***

*Writer Story Similarity: Comparison of Regression Results of Our Modified Experiment to Doshi and Hauser (2024)'s as Reported in Their Supplementary Materials, Table S11 (page 10).*

| | Our Findings | Doshi and Hauser (2024) Findings |
|---|---|---|
| Human with 1 GenAI idea | -0.493 | 0.871*** |
| | (0.5) | (0.227) |
| Human with 5 GenAI idea | -0.356 | 0.718** |
| | (0.5) | (0.240) |
| Constant | 90.588*** | 89.961*** |
| | (0.3) | (0.161) |
| Observations | 97 | 293 |
| F-Stat | 0.57 | 8.23 |
| Adj R-squared | -0.009 | 0.044 |

Note: * p < 0.05, ** p < 0.01, *** p < 0.001.

*Table 3*

*Summary Statistics of Cosine Similarity of Stories in Each Condition*

| | 0-plot (human only) | 1-plot | 5-plot |
|---|---|---|---|
| Count | 32 | 33 | 32 |
| Mean | 90.588 | 90.095 | 90.233 |
| std | 2.383 | 1.864 | 1.349 |
| min | 85.54 | 86.21 | 86.78 |
| 25% | 89.34 | 88.70 | 89.56 |
| 50% | 91.27 | 90.39 | 90.23 |
| 75% | 92.53 | 91.07 | 91.20 |
| max | 93.53 | 93.86 | 92.55 |

### 3.3.2 Mixed-effect modeling analysis

To address the statistical challenges inherent in analyzing pairwise similarity comparisons, we employed a comprehensive analytical framework combining mixed-effects modeling, Bayesian inference, and power analysis. The primary challenge stems from the non-independence of observations: each story appears in multiple pairwise comparisons, creating hierarchical dependencies that violate standard statistical assumptions.

*Mixed-Effects Linear Regression:* We fitted a mixed-effects model with experimental condition as a fixed effect and random intercepts for both stories in each pairwise comparison:

similarity ~ condition + (1|story1_id) + (1|story2_id)

This specification accounts for the fact that stories systematically differ in their baseline similarity to other stories while testing whether experimental conditions affect overall similarity patterns. The model was estimated using restricted maximum likelihood (REML) on 1,520 pairwise comparisons (496 from the 0-plot condition, 528 from the 1-plot condition, and 496 from the 5-plot condition).

The mixed-effects analysis (results summary in Table 4 and visual in Figure 2) revealed that stories in the 1-plot condition exhibited significantly lower similarity compared to the control condition ($\beta = -1.073$, 95% CI [-1.971, -0.175], $p = 0.019$), indicating greater diversity. The 5-plot condition showed a similar pattern of reduced similarity ($\beta = -0.488$, 95% CI [-1.394, 0.418], $p = 0.291$), though this difference did not reach conventional statistical significance. The estimated random effects variance was 2.76, indicating substantial between-story variation, while residual variance was 5.45.

*Story-Level Analysis as Robustness Check:* To validate these findings with a more conservative approach, we calculated each story's average similarity to all other stories within its condition, creating 97 independent observations. One-way ANOVA on these story-level averages yielded $F(2, 94) = 2.33$, $p = 0.103$. Post-hoc pairwise comparisons showed consistent patterns: 0-plot versus 1-plot (t = 1.88, p = 0.064), 0-plot versus 5-plot (t = 1.45, p = 0.151), and 1-plot versus 5-plot (t = -0.73, p = 0.470). A permutation test with 10,000 iterations confirmed these results (p = 0.113), providing non-parametric validation robust to distributional assumptions.

***Table 4***

*Mixed Linear Model Regression Results*

| | Coefficient | Standard Error | z-statistic | p-value | 95% Confidence Interval | |
|---|---|---|---|---|---|---|
| Baseline (Control: 0-plot) | 82.699 | 0.327 | 252.961 | 0.000 | 82.058 | 83.340 |
| 1-plot effect (vs. Control) | -1.073 | 0.458 | -2.341 | 0.019* | -1.971 | -0.175 |
| 5-plot effect (vs. Control) | -0.488 | 0.462 | -1.056 | 0.291 | -1.394 | 0.418 |
| Story-level variance | 2.765 | 0.212 | | | | |

Note: * p < 0.05, ** p < 0.01, *** p < 0.001.

***Figure 2***

*The frequency distribution of cosine similarity score of stories in the control(0-plot), 1-plot and 5-plot conditions in the mixed-effects analysis. The 1-plot condition exhibited significantly lower similarity compared to the control condition.*

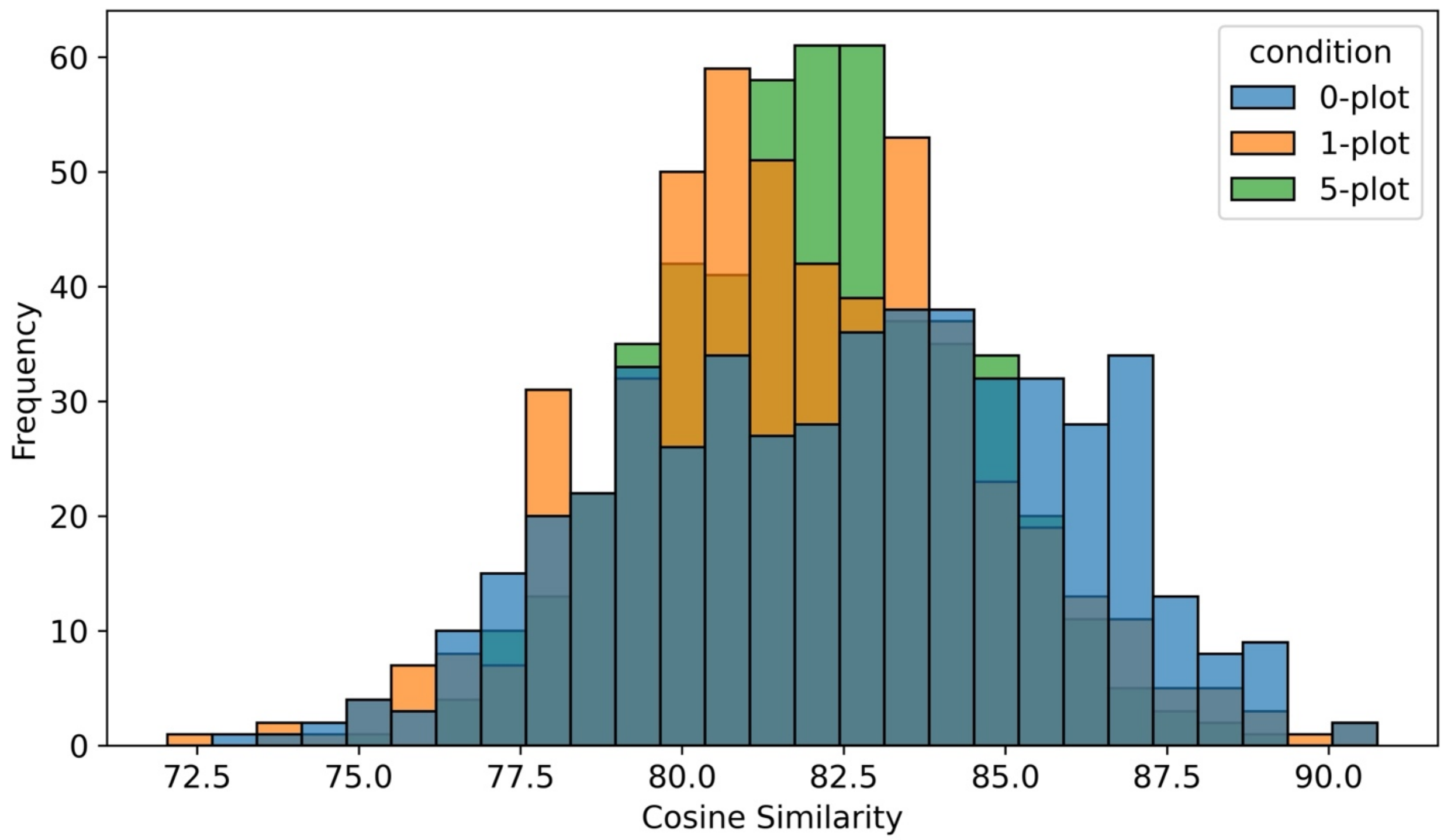

*Bayesian Analysis.* We complemented frequentist analyses with Bayesian inference to quantify evidence for equivalence between conditions. Using weakly informative priors, we fitted a Bayesian hierarchical model with 4 chains of 2,000 iterations each ($\hat{R}$ = 1.0 for all parameters, indicating convergence). The posterior distributions showed that the 1-plot condition had lower mean similarity than the control (mean difference = -0.88, 95% HDI [-1.69, -0.04]), while the 5-plot condition also showed reduced similarity (mean difference = -0.63, 95% HDI [-1.49, 0.20]).

We conducted equivalence testing using a Region of Practical Equivalence (ROPE) of ±0.5 similarity points. For the comparison between 5-plot and 1-plot conditions, 68.2% of the posterior distribution fell within the ROPE, providing moderate evidence that these two AI-assisted conditions produce practically equivalent diversity. Bayes factors comparing conditions against a null hypothesis of no difference yielded $BF_{01} = 0.55$ for 1-plot versus control (inconclusive), $BF_{01} = 1.72$ for 5-plot versus control (weak evidence for equivalence), and $BF_{01} = 3.97$ for 5-plot versus 1-plot (moderate evidence for equivalence).

*Power and Effect Size Considerations:* Post-hoc power analysis revealed that our study achieved 55.4% statistical power to detect the observed effect sizes (Cohen's f = 0.244), compared to 96.9% power for the Doshi and Hauser (2024) study with their larger sample (n = 293). The observed effect sizes were small: Cohen's d = 0.47 for 0-plot versus 1-plot, d = 0.36 for 0-plot versus 5-plot, and d = -0.18 for 1-plot versus 5-plot. Prospective power calculations indicate that detecting our observed effect at 80% power would require approximately 165 total participants (55 per condition).

The convergent evidence from mixed-effects modeling, Bayesian analysis, and permutation testing consistently demonstrates that diverse AI personas do not produce the homogenization effect observed in prior research. The 1-plot condition showed statistically significant evidence of enhanced diversity relative to the human-only baseline, while the 5-plot condition showed a similar but non-significant pattern. Importantly, the Bayesian analysis indicates moderate evidence that both AI-assisted conditions are practically equivalent to each other in terms of diversity preservation. While our study was underpowered relative to Doshi and Hauser's (2024) investigation, the small observed effect sizes and Bayesian evidence for equivalence suggest that meaningful homogenization effects are unlikely when AI inputs are intentionally diversified. These findings should be interpreted as proof-of-concept that persona-based approaches can prevent the homogenization effect, with replication in larger samples needed for definitive conclusions about diversity enhancement.

### 3.3.3 Additional linguistic difference analysis

In addition to using the above two methods, we analyzed the linguistic differences between stories assisted by AI (the 1-plot and 5-plot conditions) and stories based on humans only (the 0-plot condition) with traditional text analytics. We used the Natural Language Toolkit (NLTK) (Bird et al., 2009), a Python library for natural language processing, and its built-in corpora, particularly its English stopwords corpus, for filtering common words, and its WordNet lexical database for word categorization and semantic analysis. The text processing pipeline employed NLTK's sentence and word tokenizers for segmentation, followed by its part-of-speech tagger for identifying narrative elements such as actions (verbs), descriptions (adjectives), and character names (proper nouns), enabling systematic quantification of story elements across the dataset.

Our analysis revealed that AI-assisted stories featured greater diversity in descriptive words ($t = -2.434$, $p = 0.017$) and emotional words ($t = -2.592$, $p = 0.011$) compared to human-only stories (Table 5). There were no significant differences between AI-assisted 1-plot and 5-plot stories (Table 6). These findings suggest that while AI-assisted narratives exhibit content diversity similar to human-only stories based on cosine similarity, they demonstrate richer linguistic features, indicating that AI assistance can enhance narrative complexity and expressiveness, consistent with earlier findings that single AI plot exposure to human participant led to statistically significant more diverse stories.

***Table 5***

*Statistical Comparison Summary of Linguistic Differences Between Human-Only vs. AI-Assisted Conditions*

| Attribute | Human Mean | AI-Assisted Mean | Human Std | AI-Assisted Std | t-statistic | p-value |
|---|---|---|---|---|---|---|
| Total Words | 144.634 | 158.625 | 55.89 | 53.602 | -1.247 | 0.215 |
| Unique Words | 89.415 | 99.232 | 29.196 | 29.354 | -1.631 | 0.106 |
| Common Words % | 37.322 | 36.593 | 4.299 | 3.322 | 0.942 | 0.349 |
| Action Words | 12.951 | 14.232 | 3.619 | 3.823 | -1.667 | 0.099 |
| Descriptive Words | 15.854 | 18.321 | 5.116 | 4.794 | -2.434 | 0.017* |
| Named Characters | 0.634 | 0.946 | 0.733 | 1.227 | -1.45 | 0.15 |
| Emotional Words | 6.829 | 8.018 | 2.12 | 2.308 | -2.592 | 0.011* |

Note: * $p < 0.05$, ** $p < 0.01$, *** $p < 0.001$.

***Table 6***

*Statistical Comparison Summary of Linguistic Differences Between Two AI-Assisted Conditions*

| Attribute | 1-plot mean | 5-plot mean | 1-plot Std | 5-plot Std | t-statistic | p-value |
|---|---|---|---|---|---|---|
| Total Words | 159.963 | 157.379 | 46.944 | 59.951 | 0.179 | 0.859 |
| Unique Words | 100.815 | 97.759 | 26.458 | 32.215 | 0.386 | 0.701 |
| Common Words % | 36.27 | 36.893 | 3.482 | 3.199 | -0.698 | 0.488 |
| Action Words | 14.37 | 14.103 | 3.488 | 4.169 | 0.259 | 0.797 |
| Descriptive Words | 18.889 | 17.793 | 4.66 | 4.938 | 0.853 | 0.398 |
| Named Characters | 0.926 | 0.966 | 1.107 | 1.349 | -0.12 | 0.905 |
| Emotional Words | 7.926 | 8.103 | 2.235 | 2.41 | -0.285 | 0.777 |

Note: * $p < 0.05$, ** $p < 0.01$, *** $p < 0.001$.

# 4 Discussion

Our findings offer broad insights into the relationship between generative AI and human creativity, particularly regarding the impact of GenAI on idea diversity within collaborative settings. We will begin by discussing a relatively general conclusion about study design that stems from our study and findings, and then discuss, in subsequent subsections, more specific issues.

The proposition that GenAI reduces collective diversity, as reported by Doshi and Hauser (2024) and by others warrants critical scrutiny. A long-standing critique in communication research highlights that when *new technologies* are studied, experimental designs often default to equalizing surface-level inputs – such as time, message count, or, as in the case of Doshi and Hauser (2024), informational prompts – without sufficiently accounting for differences in how those inputs function across modalities. For example, Walther's work on computer-mediated communication demonstrated how such assumptions led to misleading conclusions in early comparisons between face-to-face and text-based interaction, particularly when asynchronous media were judged as deficient based on matched but misaligned resource (Walther et al., 2010; Walther & Tidwell, 1995). In practice, different conditions (in that case different media) require different forms of engagement to yield comparable outcomes – a principle that extends to GenAI-supported ideation.

In particular, Walther et al. (2010) draw on Rafaeli's concept of interactivity to emphasize that communication technologies vary in the degree to which one participant's contribution is contingent on another's (Rafaeli, 1988). This notion is directly relevant to the design of GenAI systems, which can operate as either static prompt generators or interactive co-creators. In the Doshi and Hauser (2024) study, the GenAI condition offered a fixed prompt to all participants, effectively eliminating the possibility for adaptation or interaction. From this perspective, the observed reduction in diversity may be less about inherent limitations of GenAI and more about the specific, non-interactive way in which it was implemented in this one particular study.

Our study speaks to this possibility by demonstrating that the relationship between GenAI and diversity is contingent on how GenAI is deployed. When GenAI outputs are themselves diverse and responsive to different narrative perspectives, they appear to amplify rather than suppress variation in human creativity. These findings align with a broader view of GenAI not as a fixed intervention but as a configurable partner whose design and use can significantly shape downstream outcomes. This view of configurability also impacted our approach to the parameters of the experiment, e.g. the number of personas used (10), the number of plots per persona (30), and the temperature used in the production of the plots (0.9), all of which were selected not as fixed or carefully derived parameters, but rather as parameters that were good enough (i.e. satisficing) to get the task done.

Building on this broader perspective, we now return to our core findings in the subsequent discussion, which show that the diversity of GenAI ideas, rather than the presence of GenAI per se, plays a central role in shaping the diversity of GenAI-inspired human-generated outputs.

## 4.1 Homogenization Effect and Multiple Pathways to Diversity

Our findings provide evidence that the homogenization effect in human-AI collaboration may be related to the manner in which GenAI was initially used to generate creative output, rather than an inherent attribute of GenAI-assisted creative work. As we demonstrate in our experiment, the homogenization observed in Doshi and Hauser (2024) stemmed from a design that focused only on augmenting creativity, and did not explicitly address diversity. Other studies (Chen & Chan, 2024; Dell'Acqua et al., 2023; Veselovsky et al., 2023) were less detailed in their description of the prompts used, so were not amenable to direct testing, but they too are likely to have created the homogenization due to not paying specific attention to the diversification of the GenAI outputs. These inadvertently created a shared creative constraint that narrowed the collective solution space, even as individuals benefitted from AI assistance. The underlying problem is simple: the studies essentially assign the same AI collaborator to every participant, whereas in pre-AI creative team work, we intuitively understand that diversity of perspectives brought by different human team members enhances collective creativity. Our persona-based approach represents just one solution among many for applying this established principle to human-AI collaboration.

Beyond personas, multiple pathways exist for diversifying AI contributions: cultural and disciplinary perspectives (embodying different viewpoints or knowledge domains), cognitive styles (analytical versus intuitive approaches), multi-model ensembles (leveraging naturally diverse AI architectures), dynamic randomization strategies (systematically varying AI outputs), and collaborative architectures (simulating diverse team roles). These and others are explored in Meincke et al. (2024). Such approaches share a common theoretical foundation – creativity emerges from intersecting different perspectives, knowledge bases, and problem-solving approaches. In human teams, this intersection occurs naturally through participant diversity; in human-AI collaboration, it must be intentionally designed into system architecture.

This finding also highlights a principle of human-AI interaction: GenAI systems tend to deliver what we explicitly request, but rarely what we implicitly assume. If we ask AI to “be creative” without specifying the need for diverse perspectives, we should not be surprised when the system optimizes for creativity metrics while inadvertently homogenizing outputs. Indeed, recent work shows that large language models converge toward narrower idea sets unless prompts explicitly encourage variety; only when users add structured guidance – such as chain-of-thought phrasing – does meaningful diversity emerge (Meincke et al., 2024). Similarly, if we prompt for “innovative solutions” without explicitly implementing varied approaches, methodologies, or frameworks, AI may converge on popular or well-represented solution patterns in its training data. A similar pattern is seen in text-to-image generation, where visual diversity closely mirrors prompt diversity: reliance on popular prompt templates produces less varied images (De Rosa Palmini & Cetinic, 2025). This pattern extends beyond creativity tasks – when we ask AI to “write persuasively” without specifying diverse rhetorical strategies, or to “analyze data” without requesting multiple analytical lenses, we often receive competent but convergent outputs. The implication is that effective human-AI collaboration requires researchers and practitioners to make their implicit assumptions about diversity, perspective, and variation explicit in their interaction design, treating AI not as an *intuitive* collaborator who can extract implicit unstated needs, but as a powerful tool that excels at delivering precisely what is specified.

Finally, we situate our findings in the context of research on the homogenization and fixation that accompany GenAI-assisted work. In controlled ideation tasks, participants who used LLM generated ideas produced ideas that were more similar to one another at the group level than ideas produced with an alternative creativity-support tool (Anderson et al., 2024). In visual design, exposure to AI-generated suggestions increased design fixation and reduced fluency, variety, and originality (Wadinambiarachchi et al., 2024). Another type of homogenization effect was demonstrated when participants who received coach-like strategy guidance from GenAI first improved their performance, but when the assistance was removed, their group-level diversity was lower in comparison with participants who received no GenAI assistance, or standard answer-giving assistance (Kumar et al., 2025). Another example of such homogenization is a natural experiment in college-admissions essays which shows a "deep homogenization" pattern: after ChatGPT's release, the lexical diversity of the essays increased, but the diversity of ideas within essays and thematic distinctiveness across essays declined (Moon et al., 2025). In conclusion, when we aspire to augment creativity with GenAI generated inputs, it is important to carefully evaluate not only the impact on the immediate creativity of the outputs, but also on the collective diversity of these outputs, the medium- to long-term impact on aided and unaided creativity of the participants, the collective diversity of the outputs, and even their psychological mindset (e.g. intrinsic motivation, Wu et al. (2025)).

## 4.2 Creativity and Diversity in Human-AI Collaboration

Recognition that the homogenization effect is methodological rather than fundamental has interesting implications for Human-AI collaboration system design. It suggests that users and organizations should avoid homogeneous deployment, build multi-perspective AI system through agentic design, and prioritize collective outcomes alongside individual performance. Our persona-based approach provides concrete evidence for this diversification principle, but the broader insight is that AI collaborators should be at least as diverse as human collaborators. As we move toward sophisticated human-AI creative partnerships, the key is recognizing that diversity is a critical tool for unlocking the full creative potential of collaborations.

The relationship between diversity and creativity is more nuanced than is often assumed, particularly in studies of collective ideation. According to the standard definition, creativity requires both originality and effectiveness (Runco & Jaeger, 2012). This definition applies equally to individuals and to groups, but its implications vary: an idea can be original and useful for an individual even if others independently produce a similar idea, while collective creativity often involves a different logic. Harvey (2014) emphasizes that extraordinary group creativity emerges not from the mere aggregation of original contributions, but from the synthesis and integration of diverse perspectives into novel combinations. From this perspective, convergent originality – when multiple individuals produce the same creative response – may still meet the criteria for individual creativity, but it can reflect a narrowing of the collective search space and limit the potential for creative synthesis. This distinction becomes especially important in the context of GenAI. Studies of large language models have shown that, in the absence of prompts that explicitly request variation, GenAI tends to generate high-probability responses that reflect dominant patterns in its training data (Meincke et al., 2024). Our findings reinforce this pattern: when GenAI ideas were uniform and unvarying, participant stories converged; when GenAI ideas reflected multiple distinct perspectives, the resulting human outputs were as diverse as those in human-only conditions. The implication is that creative quality and creative diversity are both desirable, but not synonymous – and that claims about GenAI's creative potential must attend to the diversity of GenAI's contributions, not just their originality.

What about the usage of GenAI-personas to impact diversity? Recent work on persona prompting demonstrate that personas change model behavior, but the effects are hard to predict. Adding roles to the system prompt rarely improves task accuracy and the impact varies by role, model, and task (Zheng et al., 2024). A broad audit across a large set of LLMs and several datasets shows that personas add large variance, and can shift bias and model refusal (the model refuses to provide a response to the prompt) in uneven ways (Luz de Araujo & Roth, 2025). A complementary study finds that personas grounded in psychological traits can alter theory-of-mind reasoning (Yeo et al., 2025). The results of these studies do not argue against personas; they argue for purposeful use. Our findings fit this view: when personas are treated as a controlled source of variation they can preserve, or even enhance, collective diversity, helping avoid the over-standardization seen with uniform deployment.

These considerations point to a broader lesson: as we seek to augment creativity through generative AI, we must avoid the trap of over-systematizing it. Creativity thrives on tension – between structure and surprise, constraint and divergence – and while structured techniques such as Systematic Inventive Thinking (SIT) (Goldenberg & Mazursky, 2002) show that creativity can be guided, it rarely emerges from rigid replication. GenAI systems offer powerful tools, but they do not contain a finite recipe for creativity. Rather than pursuing complete automation of creative processes, researchers and practitioners should approach creativity itself *creatively* — designing systems that invite exploration, multiplicity, and interpretive richness. In this light, the diversity-enhancing configurations we explored are not endpoints, but starting points for a more expansive understanding of human-AI creative partnerships.

## 4.3 Evidence for Diversity Enhancement

Building on our demonstration that persona-based approaches can successfully preserve collective diversity, our findings point toward an even more transformative possibility: that strategic human-AI collaboration may not merely maintain but actively enhance creative diversity beyond what humans achieve independently. This progression from diversity preservation to diversity enhancement represents a fundamental shift in how we conceptualize the potential of human-AI creative partnerships.

Our experimental results provide compelling initial evidence for this enhanced creative potential. The linguistic diversity analysis revealed that AI-assisted stories demonstrated significantly greater richness in both descriptive vocabulary ($t = -2.434$, $p = 0.017$) and emotional language ($t = -2.592$, $p = 0.011$) compared to human-only narratives. This finding is particularly noteworthy because it emerges from the same conditions that preserved semantic diversity – suggesting that our persona-based approach simultaneously maintains collective coherence while introducing linguistic sophistication that individual human creators might not achieve alone.

The pattern observed in mixed-effect modeling further supports this interpretation. The 1-plot condition showed not just preservation of diversity, but a statistically significant reduction in similarity scores relative to the human-only baseline ($\beta$ = -1.073, p = 0.019), indicating that exposure to a single diverse AI persona actually enhanced story differentiation. While the 5-plot condition showed a similar directional pattern ($\beta$ = -0.488, p = 0.291), the effect was more modest and did not reach conventional significance thresholds. The Bayesian analysis provides additional nuance: the moderate evidence for practical equivalence between the two AI-assisted conditions ($BF_{01}$ = 3.97) suggests that both 1-plot and 5-plot exposures achieve similar diversity outcomes, with the enhancement effect most pronounced and statistically detectable in the 1-plot condition.

These converging lines of evidence – linguistic enrichment, reduced semantic similarity, and Bayesian support for diversity preservation – suggest a possible mechanism: diverse AI personas may serve as creative catalysts that activate different conceptual spaces within human cognition. When participants encounter a single well-differentiated AI perspective, it may function as a productive constraint that directs their creativity into unexplored narrative territories while still allowing substantial individual variation. Multiple AI perspectives (as in the 5-plot condition) provide additional creative options but may introduce complexity that moderates the enhancement effect, resulting in outcomes practically equivalent to single-exposure conditions.

However, these interpretations must be tempered by important statistical considerations. Our study achieved only 55.4% power to detect the observed effects, and the small effect sizes (Cohen's f = 0.244) mean that we cannot conclusively distinguish between genuine enhancement and random variation. The Bayesian analysis, while providing moderate evidence against large differences between conditions, does not constitute definitive proof of equivalence. The inconclusive Bayes factors for comparisons with the control condition ($BF_{01}$ = 0.55 for 1-plot, $BF_{01}$ = 1.72 for 5-plot) indicate that our data do not strongly favor either the null hypothesis or the alternative.

This evidence has notable implications for our current practice in human-AI ideation tasks design and expands our understanding of synergy in human-GenAI collaboration. Our results suggest that humans could fully utilize GenAI's conversational and creational affordances to introduce diversity into the collaboration dynamics (Ramaul et al., 2024), treating GenAI as an agentic participant or participants in collaborative efforts. The findings indicate that diversity enhancement is possible when AI contributions are intentionally varied, though the strength and consistency of this effect requires validation in adequately powered studies with samples of approximately 165 participants or more to achieve definitive conclusions.

### 4.4 Theoretical Foundations for Creative Enhancement

This enhancement effect also aligns well with emerging theoretical insights about the representational capacity of foundation models. Advances in mechanistic interpretability suggest that large language models encode sparse, high-dimensional representations of concepts and computations, which are then projected into dense, efficient forms for practical execution (Bricken et al., 2023). This implicit sparsity enables foundation models to represent considerably more features and patterns than their explicit dimensions would suggest—essentially providing access to a vast reservoir of creative possibilities that individual humans may never encounter through conventional ideation processes.

Our persona-based approach specifically harnesses this representational richness by systematically sampling from different regions of the model's creative space. Rather than accessing only the statistically central patterns that emerge from uniform prompting, diverse personas guide the model toward exploring the periphery of its creative knowledge – regions that may contain novel combinations of cultural perspectives, narrative structures, and thematic elements that expand the creative palette available to human collaborators.

## 5 Limitations and Future Research

This study demonstrates that the trade-off between creativity and diversity is not inevitable, as might be deduced from definitive manuscript titles such as "*Generative AI* enhances individual creativity but *reduces the collective diversity of novel content*" (Doshi and Hauser, 2024). Rather, it demonstrates that the homogenization effect reported in the literature can be mitigated. Yet, the study has several limitations that point to the many directions future research could take. The statistical power of phase 2 is limited and our claims therefore rest on the consistent directional pattern across multiple analytical approaches, namely mixed-effects modeling, Bayesian analysis, permutation testing, and linguistic analysis, rather than on any single definitive test. Furthermore, our study focuses only on the mitigation of the homogenization effect. Future research should explore these limitations and expand our understanding of the unintended consequences of GenAI assisted creative work by scaling to larger and/or additional populations; using additional LLM models; testing more elaborate and more adaptive diversity controls; expanding to other contexts and languages; finding alternative methods to help human participants become more creative with GenAI; and, even achieving similar results without resorting to more computationally costly approaches as we did.

The study joins the few studies that already encourage acknowledging the unintended consequences of GenAI assisted creative work that include not only the tradeoff we discussed here, but also other forms of homogenization and fixation "caused" by GenAI. We, as well as the others (e.g. Meincke et al. (2024), Bordas et al. (2025)) remind researchers and practitioners that GenAI should not be presumed to be the primary or sole cause of such unintended consequences.

# 6 Conclusion

This research empirically challenges the assumption that human-AI creative collaboration inevitably involves a trade-off between individual enhancement and collective diversity. Through a systematic two-phase experimental design, our findings provide initial evidence that this homogenization is not intrinsic to generative AI, but may emerge from design choices that constrain variation. While our Phase 2 results are underpowered and require replication with larger samples, the consistent directional pattern across multiple analytical approaches supports this interpretation. Our persona-based framework treats AI as multiple, distinct creative collaborators rather than a singular tool. Rather than accepting reduced diversity as the price of AI-assisted creativity, the persona-based approach preserves and may enhance collective diversity while amplifying linguistic richness. Whether the individual creativity gains documented by Doshi and Hauser (2024) are maintained under diversified conditions remains an important question for future research. These findings have practical and theoretical implications. For practitioners, they provide a roadmap for designing GenAI systems that avert homogenizing group creativity. For researchers, they suggest that diversity should be treated not as a byproduct, but as a core design principle in human-GenAI creative tasks. And for the broader field of human-AI collaboration, they invite a conceptual shift**:** rather than trying to automate creativity through standardized inputs**,** it points to the futility of trying to fully automate creativity. Structured approaches like Systematic Inventive Thinking show that creativity can be guided, but not fully automated**.** As GenAI becomes a partner in more creative domains, we propose being creative about creativity itself (e.g. Meincke et al. (2024), Bordas et al. (2025)) and designing systems that encourage multiplicity, surprise, and the ongoing exploration of possibility**.**

**Declaration of generative AI and AI-assisted technologies in the writing process**: During the preparation of this work the authors used two GenAI tools, ChatGPT and Claude, in order to improve language and readability by making sentences more concise, providing feedback on the flow of the text, rephrasing and splitting long sentences, brainstorming titles, headers and ideas, and checking formatting. After using this tool/service, the authors reviewed and edited the content as needed and takes full responsibility for the content of the publication.

Supplementary Materials for

# Diverse AI Personas Can Mitigate the Homogenization Effect in Human-AI Collaborative Ideation

### Developing the ten diverse personas

To explore the impact of diversity in plot ideas on creative outputs, we developed a range of diverse personas for generating plot ideas using ChatGPT's o1preview model. The process began with an initial prompt requesting categories on which personas could differ, aiming to maximize diversity. This prompt yielded suggestions such as cultural background, literary genre preferences (excluding adventure which is the main genre of the story participants were asked to write), and thinking styles. Building on these, additional categories were generated, resulting in a comprehensive list of 14 categories, each with 4–20 classes.

The experiment utilized 14 categories for persona differentiation: Cultural Background (East Asian, South Asian, Middle Eastern, African, European, Latin American, North American, Indigenous Cultures, Pacific Islander, Caribbean); Preferred Literary Genres excluding adventure (Mystery, Science Fiction, Romance, Horror, Fantasy, Historical Fiction, Thriller, Comedy, Drama, Poetry, Non-fiction, Magical Realism, Dystopian, Biography, Satire, Mythology, Children's Literature, Young Adult, Crime, Graphic Novels); Proficiency in Using Creativity Tools (Six Thinking Hats, Mind Mapping, SCAMPER Technique, Lateral Thinking, Brainstorming, Storyboarding, Role Playing, Synectics, TRIZ Method, Random Word Technique); Thinking Styles (Analytical, Creative, Logical, Emotional, Global (Big Picture), Detail-Oriented, Divergent Thinking, Convergent Thinking, Sequential, Holistic); Technological Proficiency (Technophile (Early Adopter), Average User, Technophobe, Digital Native, Casual User, Power User); Decision-Making Style (Data-Driven, Intuitive, Consultative, Independent, Impulsive, Deliberative, Collaborative, Delegative); Learning Style (Visual, Auditory, Kinesthetic, Reading/Writing, Social (Interpersonal), Solitary (Intrapersonal)); Communication Style (Direct, Indirect, Formal, Informal, Storyteller, Listener, Assertive, Passive); Work Style (Collaborative, Independent, Structured, Flexible, Multitasker, Focused (One Task at a Time), Leader, Follower); Conflict Resolution Style (Accommodating, Avoiding, Competing, Collaborating, Compromising); Time Orientation (Past-Focused, Present-Focused, Future-Focused, Time-Fluid (Cyclical Time Perception)); Pace of Life (Fast-Paced, Moderate, Slow-Paced, Laid-Back, Urgent); Interests and Hobbies (Sports Enthusiast, Art Collector, Amateur Chef, Gamer, Musician, Gardener, Traveler, Photographer, Writer, Dancer, Hiker, Animal Lover, Collector (Stamps, Coins, etc.), Volunteer Worker, Yoga Practitioner, Film Buff, Astronomy Enthusiast, DIY Crafter, Language Learner, Blogger); and Social Role (Caregiver, Activist, Entrepreneur, Scholar, Mentor, Student, Artist, Parent, Community Leader, Explorer). Each category contained at least four "classes," with some having up to 20, designed to ensure broad diversity without directly impacting creativity. This diverse set of traits was then used to generate detailed persona descriptions, which served as the foundation for developing plot ideas. This method ensured that the personas—and the plot ideas derived from them—captured a broad spectrum of diversity, serving as a foundation for testing the relationship between input diversity and creative outputs in human-GenAI collaborations.

**Prompting the personas for plot ideas**

These personas were incorporated into an assistant on the Azure OpenAI platform using gpt-4o version 2024-05-13, temperature=0.9, top P = 1. The prompt was:

---

You are an AI assistant that simulates different personas who write story plots. You are provided with a description of the persona, and the number of story plots that persona is asked to write.

For each of the plot ideas, generate an exactly three-sentence plot idea for a short story that is an adventure on another planet.

Important Instructions:

The plot should be understandable by an adult with 12 years of education.

Do not include any references to the persona's characteristics, traits, or theoretical constructs in the plot idea.

Read the persona's description to understand their style. Write each plot idea in the style of the persona, reflecting their unique perspective and voice.

The persona is not the protagonist of the story, and the protagonist does not have the traits of the persona.

The plot ideas should stand on their own, without any mention of the personas' descriptions.

The output should include only the plot idea, without titles, numbering, or anything else.

Plan the plot idea in exactly three sentences, ensuring each sentence adds to the story.

The plots should not be similar to a previous plot

The format of the output is a csv file

---

The prompt that was used in the "advisor": "Create thirty-one plot ideas of exactly three sentences each that [persona name] would write". The first thirty were selected and saved. ChatGPT was used to verify the story plots are not essentially similar, after several iterations that resulted in some plots being essentially the same. This phenomenon was resolved when the temperature was raised from 0.7 to 0.9.

Below are the descriptions of the personas that were used, followed by two examples of the story plots produced by each. A full list of the plots is available from the corresponding author.

**Persona 1: Amina**

Meet Amina, a Middle Eastern woman who adores historical fiction. She's proficient in mind mapping as a creativity tool and has a global (big picture) thinking style. As a technophile (early adopter), she eagerly adopts new gadgets and apps. Amina makes decisions intuitively, trusting her gut feelings. She prefers the visual learning style and communicates in a formal manner. Her collaborative work style makes her a team player who values everyone's input. In conflicts, she aims for compromising solutions. Being future-focused, she sets long-term goals and plans meticulously. She leads a fast-paced life, juggling multiple projects. Her hobbies include being a language learner and an art collector. She sees herself as an entrepreneur, always looking for the next big opportunity.

Sample plots:

| A brave archaeologist discovers an ancient map leading to a lost civilization on Mars. Determined to uncover its secrets, she joins a team of scientists and adventurers, embarking on a perilous journey through the Martian landscape. As they navigate treacherous terrain and unravel the mysteries of the past, they find clues that connect the Martian civilization to the origins of humanity. |
|---|
| In the distant future, an interstellar historian uncovers a hidden portal to a forgotten planet. The historian and his team set out to explore the planet, uncovering remnants of an advanced alien society. As they piece together the alien history, they stumble upon advanced technology that could change the course of human civilization. |

**Persona 2: Liam**

Liam is a European man passionate about science fiction. He utilizes lateral thinking to generate innovative ideas and has a logical thinking style. As a technophobe, he prefers minimal interaction with technology. Liam is data-driven in his decision-making, relying heavily on facts and statistics. He learns best through reading and writing and communicates directly. His independent work style suits his preference for working alone. In conflicts, he tends to avoid confrontation. Liam is past-focused, often reflecting on history for lessons. He enjoys a moderate pace of life. His hobbies include being an amateur chef and a gardener. He identifies as a scholar, constantly seeking knowledge.

Sample plots:

| After a routine geological survey, Dr. Evelyn discovers an anomaly that corresponds with ancient texts describing a hidden city on Planet Delta. She embarks on a solo expedition, relying on her knowledge rather than technology, to uncover the truth. What she finds not only rewrites the planet's history but also challenges her understanding of human civilization. |
| --- |
| While exploring the uncharted forests of Planet Zephyr, archaeologist Alex stumbles upon ruins that eerily resemble those of ancient Earth cultures. As he digs deeper, he realizes that these ruins hold the key to understanding a lost chapter of Earth's history. His journey becomes a race against time to protect the site from off-world treasure hunters. |

**Persona 3: Sofia**

Sofia hails from a Latin American background and loves magical realism. She is adept at storyboarding and thinks in a creative manner. A digital native, she seamlessly integrates technology into her life. Sofia's decision-making is impulsive, following her instincts. She is an auditory learner and communicates as a storyteller, captivating her audience. Her flexible work style allows her to adapt quickly to changes. In conflicts, she prefers collaborating to find win-win solutions. Being present-focused, she lives in the moment. Sofia leads a laid-back lifestyle, taking things as they come. She enjoys dancing and is an animal lover. She sees herself as an artist, expressing herself through various mediums.

Sample plots:

| An archaeologist discovers an ancient alien temple on a distant planet, revealing records of a lost civilization's history. Through their meticulous study, they uncover the secrets of the planet's past and the reason for its inhabitants' mysterious disappearance. As they piece together the data, they inadvertently activate a dormant mechanism that threatens to destroy the planet, forcing them to find a solution using only their knowledge and logic. |
| --- |
| A spaceship crash-lands on a barren planet, and the crew must find a way to survive in an environment devoid of advanced technology. They discover a series of underground caves containing untapped resources and remnants of an ancient species. By studying the historical artifacts and adapting agricultural techniques, they manage to sustain themselves and eventually re-establish contact with their home world. |

**Persona 4: Kenji**

Kenji is from an East Asian culture and enjoys mystery novels. He uses the Six Thinking Hats method for creativity and has a detail-oriented thinking style. As an average user of technology, he's comfortable but not overly reliant on it. Kenji makes deliberative decisions, weighing all options carefully. A kinesthetic learner, he learns best through hands-on experience. He communicates indirectly, valuing harmony over bluntness. His structured work style helps him stay organized. In conflicts, he is accommodating, often putting others' needs first. Kenji is time-fluid, perceiving time in a cyclical manner. He maintains a slow-paced life, enjoying each moment. His hobbies include being a gardener and a collector of rare coins. He identifies as a mentor, guiding others in his community.

Sample plots:

| A group of botanists is sent to a distant planet with the mission to catalog alien plant species. When one of the scientists goes missing, their colleagues discover a hidden underground network of bioluminescent flora that holds the key to their friend's disappearance. As they navigate the mesmerizing but perilous jungle, they must solve the mystery before the planet's wildlife catches up to them. |
|---|
| An interstellar treasure hunter lands on a red-tinged desert planet in search of an ancient artifact said to control the sands. When they find the artifact, they must decipher a series of cryptic puzzles left by an extinct civilization. In the end, the true treasure is the uncovering of the civilization's lost history, which could transform their own world. |

**Persona 5: Fatima**

Fatima comes from an African background and is fond of poetry. She excels at brainstorming and has an emotional thinking style. As a casual user of technology, she uses it when necessary. Fatima makes decisions consultatively, seeking others' opinions. A social (interpersonal) learner, she thrives in group settings. She communicates informally, making others feel at ease. Her multitasking work style enables her to handle several responsibilities. In conflicts, she aims for collaborating resolutions. Being present-focused, she cherishes the here and now. Fatima enjoys a moderate pace of life. Her hobbies include volunteer work and being a yoga practitioner. She sees herself as a caregiver, dedicated to helping others.

Sample plots:

| A group of intrepid explorers land on the lush, verdant planet of Veridia, only to discover its plant life communicates through poetry. As they bond with the vegetation, they uncover a hidden message warning of an impending cosmic storm. With time ticking away, they must collaborate with the planet's poetic guardians to shield Veridia from destruction. |
| --- |
| An adventurous team of scientists arrives on the shimmering ice planet, Glacia, inhabited by sentient crystalline beings. They must learn the unique dance-like language of the crystals to request their help in finding a cure for a deadly Earth virus. Amid this delicate dance of understanding, they forge an unbreakable bond with their sparkling hosts. |

**Persona 6: Carlos**

Carlos is from the Caribbean and loves comedy. He uses the random word technique for creativity and has a divergent thinking style. As a technophile (early adopter), he's always up-to-date with the latest tech. Carlos is impulsive in decision-making, often acting on a whim. An auditory learner, he enjoys listening to lectures and discussions. He communicates informally, often cracking jokes. His flexible work style allows him to adapt to new situations. In conflicts, he tends to compete to have his way. Carlos is future-focused, always thinking about what's next. He leads a fast-paced life, embracing the hustle. His hobbies include being a musician and a gamer. He identifies as an explorer, always seeking new experiences.

Sample plots:

| Thomas and his trusty robot, Bob, crash land on a vibrant planet made of jelly and are surrounded by sentient gummy bears. They must find a way to communicate with the gummy bears to rebuild their spaceship while avoiding being squished by giant marshmallow storms. Hilarity ensues when Bob short circuits from overloads of candy data, forcing Thomas to teach him how to dance the gummy dance to survive. |
| --- |
| A group of space tourists end up on a planet where all buildings are made of inflatable materials. When a freak helium storm hits, they have to chase down their floating hotel before it drifts away into the cosmos, all while bouncing like lunatics on air-filled streets. They discover that the planet's only way of communication is through inflatable balloon animals, leading to comical misunderstandings and unexpected friendships. |

**Persona 7: Elena**

Elena, from a North American background, is passionate about graphic novels. She is skilled in role-playing as a creativity tool and has a holistic thinking style. A power user of technology, she utilizes advanced features and software. Elena makes decisions independently, trusting her judgment. As a visual learner, she prefers diagrams and illustrations. She communicates directly, getting straight to the point. Her focused (one task at a time) work style ensures thoroughness. In conflicts, she seeks compromising solutions. Being future-focused, she plans ahead extensively. Elena enjoys a fast-paced lifestyle. Her hobbies include photography and being a film buff. She sees herself as an entrepreneur, aiming to start her own business.

Sample plots:

| Riley, a fearless astronaut, embarks on a solo mission to the planet Zyxalia to investigate a mysterious signal. Upon arrival, she discovers an ancient alien ruin teeming with advanced technology that holds the key to humanity's future. As the planet's unstable environment begins to collapse, Riley must race against time to decipher the alien tech and escape before being trapped forever. |
| --- |
| Mara, an intergalactic archaeologist, finds herself on the icy planet of Thalos where she uncovers a hidden civilization beneath the ice. As she delves deeper into the frozen tunnels, she encounters an intelligent alien species in cryogenic sleep. Awakening them, Mara realizes they have the knowledge to save their dying sun but must navigate complex cultural differences to form an alliance. |

**Persona 8: Raj**

Raj is from a South Asian culture and enjoys mythology. He uses the SCAMPER technique for creativity and has a logical thinking style. As a technophobe, he prefers traditional methods over digital ones. Raj is deliberative in decision-making, analyzing all aspects. A reading/writing learner, he absorbs information through texts. He communicates in a formal manner, respecting conventions. His structured work style helps maintain order. In conflicts, he prefers avoiding to maintain peace. Raj is past-focused, valuing traditions and history. He leads a slow-paced life, enjoying simplicity. His hobbies include being a writer and a volunteer worker. He identifies as a scholar, dedicated to lifelong learning.

Sample plots:

| On the vibrant planet of Nidar, an elder named Arjin, known for his wisdom from ancient scrolls, leads a group of explorers to find a hidden temple believed to hold the secret to eternal peace. Along their journey through treacherous landscapes, they encounter mythical creatures and ancient traps. Their faith in age-old traditions and Arjin's knowledge ultimately guide them to the temple, where they uncover a long-lost civilization's peaceful way of life. |
|---|
| In the ancient forests of Zyndor, a group of young knights embarks on a quest to retrieve a sacred artifact rumored to restore balance to their war-torn world. Guided by the stars and ancient legends, they traverse mystical terrains filled with enchanted flora and fauna. The artifact is guarded by a wise dragon, who tests their virtue before bestowing upon them the power to heal their planet. |

**Persona 9: Isabella**

Isabella hails from a European background and loves romance novels. She's proficient in synectics and has an emotional thinking style. A digital native, she seamlessly navigates technology. Isabella makes intuitive decisions based on feelings. A kinesthetic learner, she engages physically with her environment. She communicates as a storyteller, captivating her audience. Her collaborative work style makes her a great team member. In conflicts, she is accommodating, often yielding to others. Being present-focused, she enjoys the moment. Isabella leads a moderate pace of life. Her hobbies include dancing and being an animal lover. She sees herself as an artist, expressing herself creatively.

Sample plots:

| On the enigmatic planet of Elysium, Maya, an adventurous astronaut, stumbles upon a hidden society where emotions are expressed through vivid light displays. As she learns their luminous language, she falls for the captivating leader, Lior, whose emotions shine like a kaleidoscope. Together, they must navigate treacherous terrain to save his society from an impending darkness, uniting their worlds in a radiant embrace. |
|---|
| Adrift on the serene oceans of Aqua-Luna, marine biologist Aria discovers a lost civilization of sentient aquatic beings. She forms a deep connection with Kai, one of the leaders, who teaches her the art of underwater dance and communication. When a menacing storm threatens their worlds, Aria and Kai must find a way to protect their newfound love and his ancient city. |

**Persona 10: Thomas**

Thomas is from an indigenous culture and is interested in dystopian literature. He uses the TRIZ method for creativity and has a global (big picture) thinking style. As an average user of technology, he's comfortable but not enthusiastic about it. Thomas makes data-driven decisions, relying on evidence. A visual learner, he prefers images and charts. He communicates indirectly, valuing subtlety. His independent work style suits his self-reliant nature. In conflicts, he seeks collaborating solutions. Being time-fluid, he sees time as a continuous loop. He leads a laid-back life, avoiding unnecessary stress. His hobbies include hiking and being an astronomy enthusiast. He identifies as a community leader, working towards the betterment of his people.

Sample plots:

| |
|---|
| An astronaut, exploring a new planet, stumbles upon an ancient, thriving civilization, seamlessly blending nature with advanced technology, their culture rooted in community and environmental stewardship. As she integrates with them, she discovers that their peaceful existence is threatened by an impending cosmic disaster. Together, they must find a solution that preserves their way of life while addressing the environmental crisis they face. |
| A nomadic tribe on a distant planet has lived in harmony with their surroundings for centuries, guided by the wisdom of their ancestors etched in sacred stones. When a mega-corporation begins mining the planet for resources, the tribe must find a way to protect their home and heritage. With the help of an off-world ally, they uncover hidden knowledge within the stones that could save their planet from exploitation. |

We also used the prompt in Doshi and Hauser (2024) to generate 30 AI story plots to be used as benchmark comparison with the 300 story plots generated by the 10 personas. The prompt used was: "write a three-sentence summary of a story about an adventure on a different planet." We ran the prompt 30 times on the Azure OpenAI chat playground using GPT-4, to emulate running this prompt on the OpenAI API used by Doshi & Hauser (2024). In total 30 plot ideas were collected.

Sample plots:

| |
|---|
| A group of explorers lands on Zephyra, a distant, wind-swept planet teeming with bioluminescent forests and floating rock formations. While searching for a rare energy source, they encounter intelligent alien beings who communicate through music and warn them of a lurking danger beneath the surface. As the explorers race to uncover Zephyra's secrets, they must forge an alliance with the aliens to survive an ancient, planet-wide catastrophe. |
| A team of scientists embarks on an expedition to Lyvora, a vibrant alien planet covered in crystalline jungles and shifting oceans of liquid light. While studying the planet's mysterious energy fields, they encounter an ancient civilization that has been dormant for millennia but is now awakening. As tensions rise between discovery and survival, |

the explorers must unravel Lyvora's secrets to prevent a devastating chain reaction that could destroy the planet—and themselves.

## The intra- and inter-persona similarity

### Table S1

(all cosine similarity scores are scaled to 100 in the tables)

| | Intra-Personal Similarity | | | | | Inter-Persona Similarity | | | | | distinct |
|---|---|---|---|---|---|---|---|---|---|---|---|
| | **mean** | **std** | **median** | **min** | **max** | **mean** | **std** | **median** | **min** | **max** | |
| Persona 1 | 87.65 | 2.52 | 87.48 | 81.60 | 100.00 | 84.38 | 3.06 | 84.54 | 73.80 | 95.52 | 3.27 |
| Persona 2 | 87.65 | 2.38 | 87.18 | 82.73 | 100.00 | 83.96 | 2.93 | 83.97 | 74.74 | 98.54 | 3.69 |
| Persona 3 | 87.42 | 2.55 | 86.87 | 82.91 | 96.52 | 84.03 | 2.95 | 84.02 | 75.54 | 98.82 | 3.40 |
| Persona 4 | 87.74 | 3.07 | 87.07 | 82.95 | 100.00 | 84.59 | 3.28 | 84.63 | 74.94 | 99.72 | 3.15 |
| Persona 5 | 88.02 | 3.15 | 87.23 | 82.24 | 100.00 | 83.35 | 3.30 | 83.29 | 74.60 | 99.72 | 4.67 |
| Persona 6 | 87.42 | 2.72 | 87.61 | 79.92 | 95.01 | 79.73 | 1.75 | 79.76 | 73.24 | 85.94 | 7.69 |
| Persona 7 | 85.17 | 2.56 | 85.19 | 79.32 | 97.78 | 82.69 | 2.76 | 82.76 | 70.43 | 92.26 | 2.48 |
| Persona 8 | 86.46 | 3.15 | 86.39 | 78.47 | 100.00 | 81.61 | 2.52 | 81.74 | 70.43 | 93.20 | 4.84 |
| Persona 9 | 86.47 | 4.09 | 85.13 | 81.74 | 100.00 | 81.86 | 2.37 | 81.72 | 74.54 | 91.11 | 4.61 |
| Persona 10 | 90.11 | 2.73 | 89.67 | 85.12 | 100.00 | 83.06 | 2.42 | 83.23 | 74.15 | 93.20 | 7.05 |
| D&H simulation* | 90.99 | 2.78 | 91.24 | 81.08 | 96.93 | * | * | * | * | * | * |

*The intra-persona statistics for 30 AI plots generated by the prompt reported in Doshi and Hauser (2024).

Figure S1

**Figure S1** presents individual persona profiles, showing how each persona's intra-similarity compares to its average inter-persona similarity with all other personas. These individual profiles reveal the substantial variation in persona effectiveness while confirming that all personas achieve positive distinctiveness.

**Figures S1: Heat-map of similarity between personas (scale to 100)**

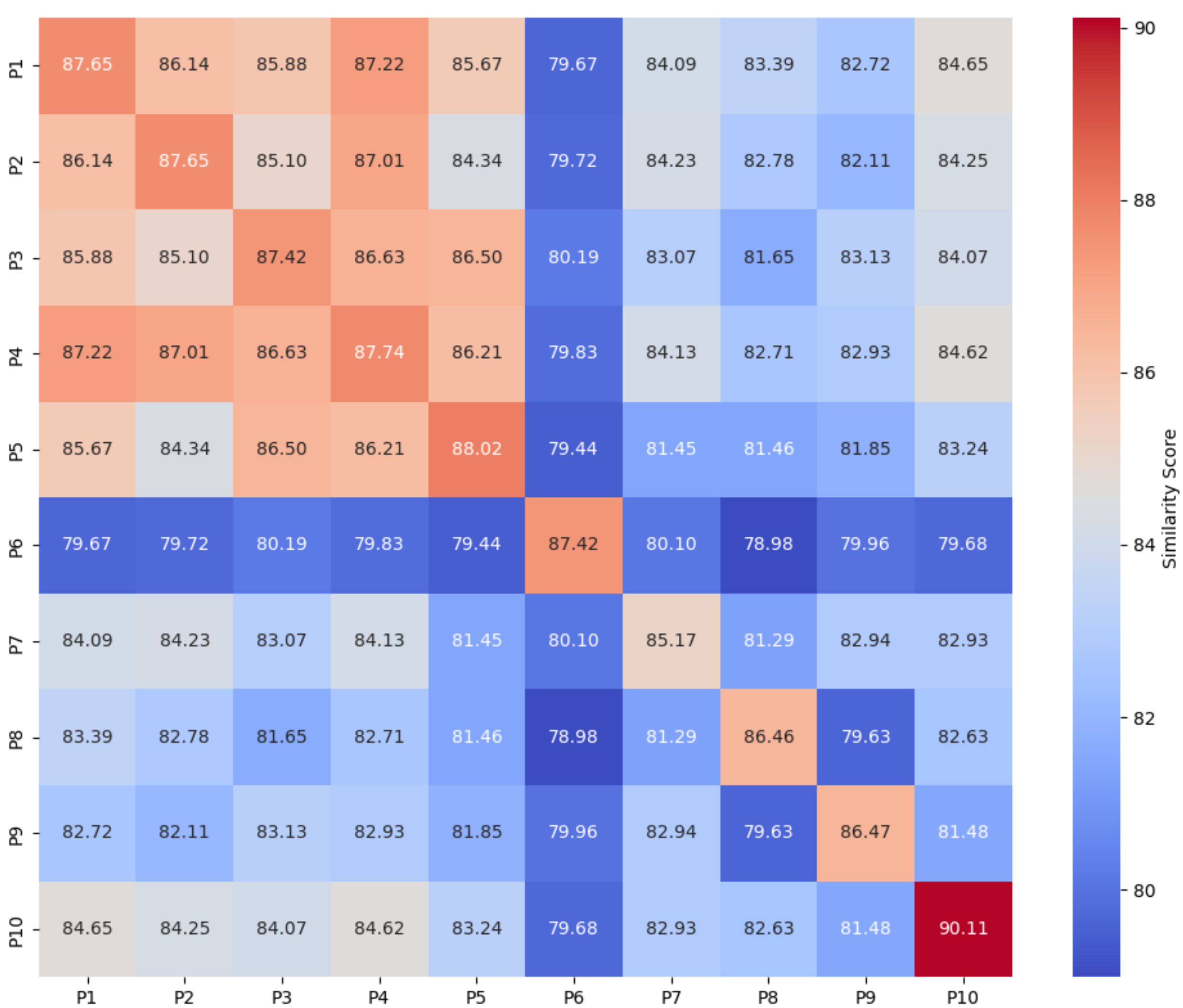

Figure S2

**Figure S2 compares the distribution of intra-persona versus inter-persona similarity scores. The graph shows that plots generated by the same persona (intra-persona, blue box) have significantly higher similarity scores (mean 87.41) compared to plots generated by different personas (inter-persona, red box, mean 82.93). The 4.49-point difference with a p-value of 0.00e+00 indicates statistically significant persona differentiation, indicating each AI persona generates distinctly consistent content compared to other personas.**

**Figure S2: Box plot comparison of** intra- and inter-persona **similarity scores distribution (scaled to 100)**

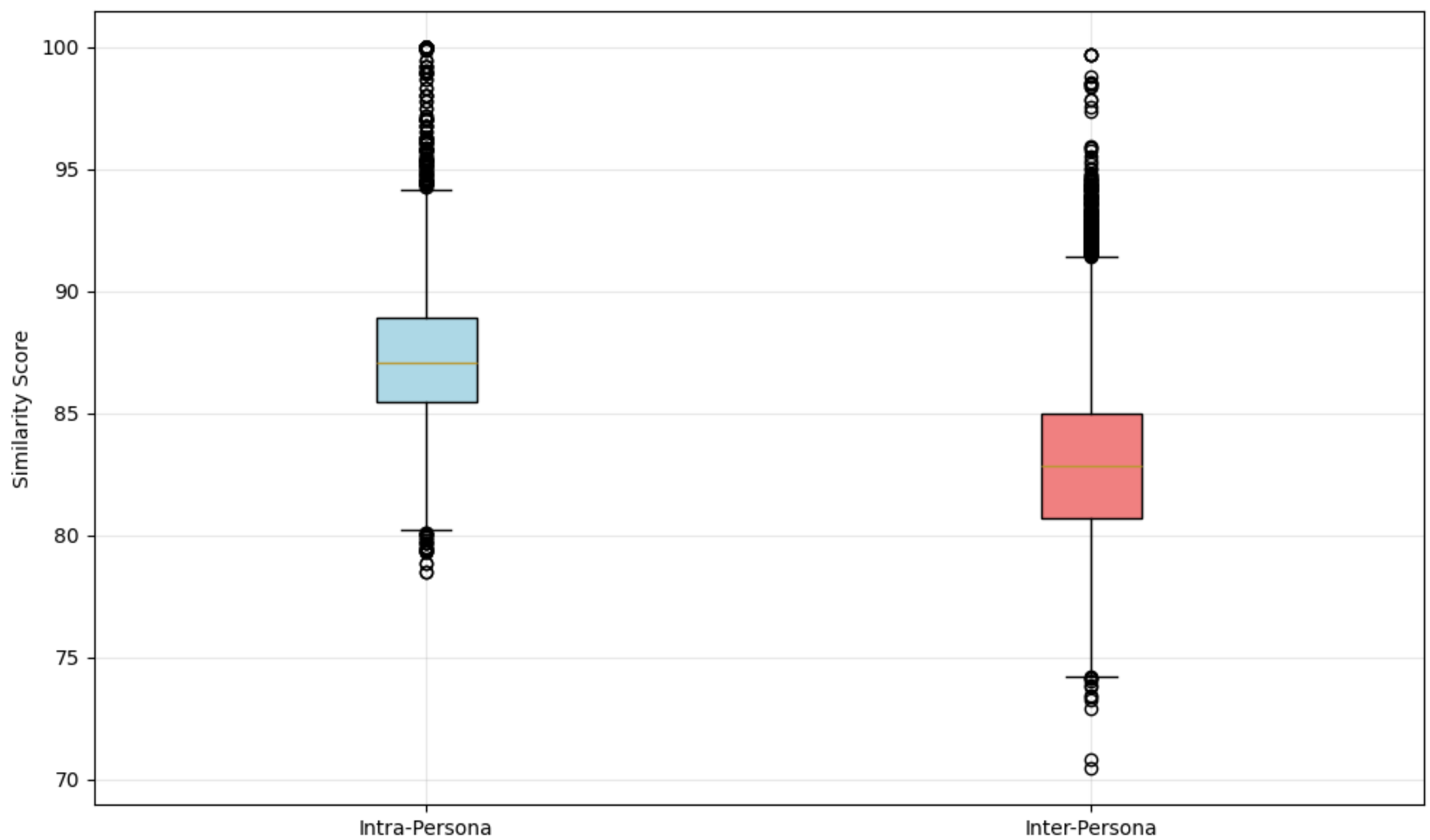

Figure S3

**Figure S3 presents** the combined normalized density plot of intra- and inter-persona similarity score distributions. It overlays the normalized density curves for intra-persona and inter-persona similarities on the same plot for direct comparison. Greater separation between curves indicates stronger persona distinctiveness.

**Figure S3:** Combined normalized density plot of intra- and inter-persona similarity score distributions

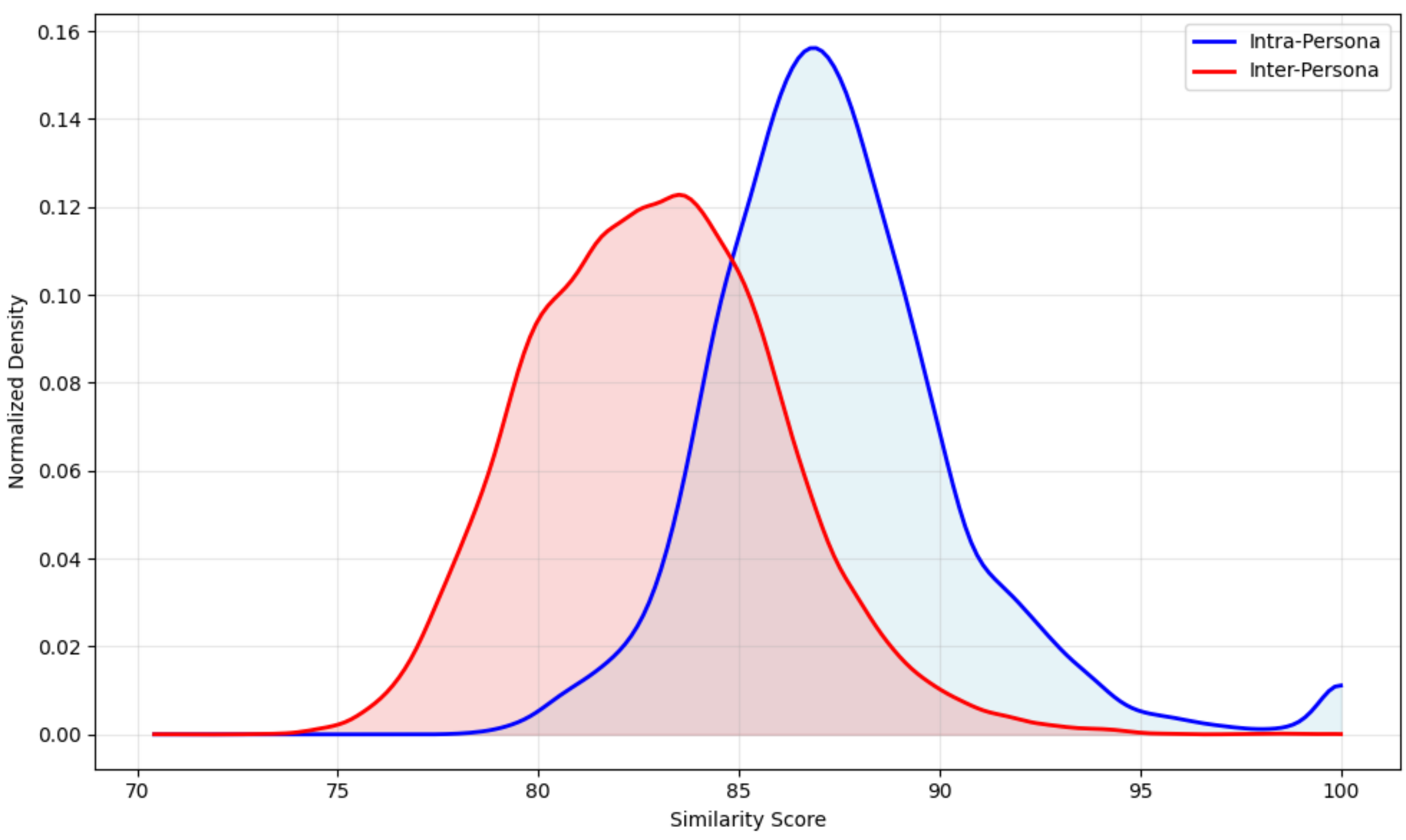

**Story writing by participants**

100 participants were recruited on Prolific, using the same recruitment criteria used by Doshi and Hauser (2024), and, using the same protocol. Each participant was randomly assigned to one of the three conditions: 0-plot (write the story with no AI assistance), 1-plot, receiving the option to get one GenAI generated plot, and 5-plot, receiving the option to get up to five GenAI generated plots. In line with Doshi and Hauser, participants could choose whether to ask for plots or not, and were not required to use the plot ideas even if they requested them. Below are the screen shots of the experiment.

## Opening page

This page describes the study you will participate in and outlines your rights as a study participant.

In this study, you will be asked to perform a short task of providing ten words based on criteria that will be provided to you, and then to write a short eight-sentence story under one of several conditions, and to answer a few follow-up questions. To ensure unbiased responses, the full details and purpose of the study will be shared with you after the study's completion. You will be debriefed at that time.

You will receive 3 pounds (UK) for your participation, which is expected to take about 15-20 minutes. If at any point you feel uncomfortable, you have the right to withdraw from the study without penalty. Should you choose to withdraw, please note that the completion code for Prolific will only be provided if you complete the study.

Your participation is entirely anonymous, and we will not collect any identifying information about you. Only non-identifiable data will be collected and stored for research purposes. There are no known or expected risks to participating in this study beyond those encountered in everyday life.

At the conclusion of the study, you will receive a completion code for Prolific, and the researchers' contact information will be provided so that you can reach out with any questions or concerns.

By clicking "continue," you confirm that you are over 18 and consent to participate in the study.

○ Continue

○ Exit the experiment

### Human only (0-plot) condition

We would like you to write a story about an adventure on another planet. You can write about anything you like. The story must be **exactly eight sentences long** and it needs to be written in English and appropriate for a **teenage and young adult audience** (approximately 15 to 24 years of age).

Please write your story here

### Verification question that followed the 0-plot condition

Please tell us whether you used ChatGPT or a similar generative AI tool to inspire your story? (Please answer truthfully: your truthful answer will help us with our research. Your answer will NOT affect your payment.)

- ◯ Yes
- ◯ No
- ◯ Other

### 1-plot condition before asking for a plot idea

We would like you to write a story about an adventure on another planet. You can write about anything you like. The story must be **exactly eight sentences long** and it needs to be written in English and appropriate for a **teenage and young adult audience** (approximately 15 to 24 years of age).

In order to assist you, we have provided you access to AI assistance that, if you wish, will come up with a starting point for your story by clicking on the "Suggest Story Idea" button below. The response will be presented below. **You are free to use or disregard any element of the AI assistant's story idea, or start over with your own idea**.

Click the button to receive a story idea:

Suggest Story Idea

Please write your story here

## 1-plot condition after asking for a plot idea

We would like you to write a story about an adventure on another planet. You can write about anything you like. The story must be **exactly eight sentences long** and it needs to be written in English and appropriate for a **teenage and young adult audience** (approximately 15 to 24 years of age).

In order to assist you, we have provided you access to AI assistance that, if you wish, will come up with a starting point for your story by clicking on the "Suggest Story Idea" button below. The response will be presented below. **You are free to use or disregard any element of the AI assistant's story idea, or start over with your own idea**.

Click the button to receive a story idea:

Suggest Story Idea

While exploring a remote alien planet, a geologist finds a vast network of crystal caves that store historical data in their formations. Analyzing the crystals' patterns, they uncover the planet's geological history and the events that shaped its surface. This information helps them predict and prepare for an imminent volcanic eruption that could threaten their mission.

Please write your story here

### 5-plots condition before asking for plot ideas

We would like you to write a story about an adventure on another planet. You can write about anything you like. The story must be **exactly eight sentences long** and it needs to be written in English and **appropriate for a teenage and young adult audience** (approximately 15 to 24 years of age).

In order to assist you, we have provided you access to AI assistance that, if you wish, will come up with a starting point for your story by clicking on the five "Idea" buttons below. Clicking on each button will present to you a story idea. The response will be presented below. You can get a new idea by clicking on a green button, or go back to an idea you already received by clicking on a yellow button. You are free to use or disregard any element of the AI assistant's story ideas, or start over with your own idea.

Idea 1 Idea 2 Idea 3 Idea 4 Idea 5

Please write your story here

## 5-plots condition after asking for a plot idea

We would like you to write a story about an adventure on another planet. You can write about anything you like. The story must be **exactly eight sentences long** and it needs to be written in English and **appropriate for a teenage and young adult audience** (approximately 15 to 24 years of age).

In order to assist you, we have provided you access to AI assistance that, if you wish, will come up with a starting point for your story by clicking on the five "Idea" buttons below. Clicking on each button will present to you a story idea. The response will be presented below. You can get a new idea by clicking on a green button, or go back to an idea you already received by clicking on a yellow button. You are free to use or disregard any element of the AI assistant's story ideas, or start over with your own idea.

Idea 1 | Idea 2 | Idea 3 | Idea 4 | Idea 5

An explorer embarks on a journey to a planet known for its floating islands that change positions with the wind. On one of these islands, they discover an ancient temple with puzzles guarding a long-lost secret. They must solve these puzzles to unlock the temples secret and harness the knowledge to stabilize the planets erratic climate.

Please write your story here

## 5-plots condition after asking for another plot idea

We would like you to write a story about an adventure on another planet. You can write about anything you like. The story must be **exactly eight sentences long** and it needs to be written in English and **appropriate for a teenage and young adult audience** (approximately 15 to 24 years of age).

In order to assist you, we have provided you access to AI assistance that, if you wish, will come up with a starting point for your story by clicking on the five "Idea" buttons below. Clicking on each button will present to you a story idea. The response will be presented below. You can get a new idea by clicking on a green button, or go back to an idea you already received by clicking on a yellow button. You are free to use or disregard any element of the AI assistant's story ideas, or start over with your own idea.

Idea 1 Idea 2 Idea 3 Idea 4 Idea 5

On the lush planet of Oasis, a group of adventurers encounters a civilization that lives in balance with its abundant natural resources. They must learn the principles of sustainable living to help Earth's strained environment. Through collaboration and shared practices, they bring hope for a greener future.

Please write your story here

### Calculating story diversity

After removing three 0-plot stories in which participants reported using outside assistance by GenAI, the 97 remaining stories were analyzed using OpenAI's API to calculate their embeddings using model "text-embedding-ada-002" with 1536 default embedding vector length. A regression was performed to detect differences between the story diversities in the three conditions (Table 2), and mean diversities were calculated for each of the three conditions (Table 3). We also used the mixed-effect modeling analysis (results in Table 4) to detect the differences. Additional statistical analysis results by using this method are presented below. Finally, we used traditional linguistic analysis with the Natural Language Toolkit (NLTK), results are presented in Tables 5 and 6.

### Table S2

Table S2: Bayesian Analysis Results

Bayesian Model Summary

| | mean | sd | hdi_2.5% | hdi_97.5% |
|---|---|---|---|---|
| Intercept | 82.616 | 0.3 | 82 | 83.197 |
| condition[1-plot] | -0.882 | 0.419 | -1.7 | -0.042 |
| condition[5-plot] | -0.625 | 0.433 | -1.5 | 0.203 |
| sigma | 1.7 | 0.126 | 1.46 | 1.956 |

Note: All parameters showed excellent convergence ($\hat{R} = 1.00$) and adequate effective sample sizes (ESS > 6,000).

Equivalence Testing

Successfully extracted 8000 posterior samples for each comparison

| 1-plot vs 0-plot: | 5-plot vs 0-plot: |
|---|---|
| Mean difference: -0.882<br>95% HDI: [-1.686, -0.042]<br>P(effect in ROPE): 0.181<br>P(negative effect): 0.818<br>P(positive effect): 0.001<br>Interpretation: Inconclusive evidence | Mean difference: -0.625<br>95% HDI: [-1.486, 0.203]<br>P(effect in ROPE): 0.375<br>P(negative effect): 0.620<br>P(positive effect): 0.006<br>Interpretation: Inconclusive evidence |

5-plot vs 1-plot:

Mean difference: 0.258
95% HDI: [-0.594, 1.080]
P(effect in ROPE): 0.682
P(negative effect): 0.040
P(positive effect): 0.279
Interpretation: Inconclusive evidence

<table>
<tr>
<td rowspan="2">Bayes Factors</td>
<td>1-plot vs 0-plot:<br>$BF_{01}$ (null vs alternative): 0.548<br>$BF_{10}$ (alternative vs null): 1.826<br>Interpretation: Inconclusive evidence</td>
<td>5-plot vs 0-plot:<br>$BF_{01}$ (null vs alternative): 1.724<br>$BF_{10}$ (alternative vs null): 0.580<br>Interpretation: Weak evidence for no difference</td>
</tr>
<tr>
<td colspan="2">5-plot vs 1-plot:<br>$BF_{01}$ (null vs alternative): 3.968<br>$BF_{10}$ (alternative vs null): 0.252<br>Interpretation: Moderate evidence for no difference</td>
</tr>
</table>

## Table S3

Table S3: Power Analysis Results

<table>
<tr>
<td rowspan="4">Intent-to-Treat Analysis</td>
<td>Originally Assigned Conditions</td>
<td>1-plot condition (81.74) < control (82.62) = MORE diverse<br>5-plot condition (82.00) < control (82.62) = MORE diverse</td>
</tr>
<tr>
<td>Observed Effect Size</td>
<td>0-plot vs 1-plot: Cohen's d = 0.467<br>0-plot vs 5-plot: Cohen's d = 0.363<br>1-plot vs 5-plot: Cohen's d = -0.180</td>
</tr>
<tr>
<td>Effect Size Interpretation</td>
<td>0-plot vs 1-plot: small effect<br>0-plot vs 5-plot: small effect<br>1-plot vs 5-plot: negligible effect</td>
</tr>
<tr>
<td>Overall ANOVA effect size</td>
<td>Observed F = 2.720<br>Eta-squared = 0.0564<br>Cohen's f = 0.244 (small)</td>
</tr>
<tr>
<td rowspan="2">Post-hoc Power Analysis</td>
<td>Achieved Power (n=97)</td>
<td>0.554 (55.4%)</td>
</tr>
<tr>
<td>Prospective Power Analysis</td>
<td>Sample Size Requirements for ANOVA with 3 Groups (α = 0.05)
<table>
<tr><td>Effect Size (Cohen's f)</td><td>80% Power</td><td>90% Power</td><td>95% Power</td></tr>
<tr><td>Medium (f = 0.40)</td><td>64</td><td>83</td><td>100</td></tr>
<tr><td>Observed (f = 0.24)</td><td>165</td><td>215</td><td>262</td></tr>
<tr><td>Small (f = 0.25)</td><td>158</td><td>206</td><td>251</td></tr>
</table>
</td>
</tr>
</table>